\documentclass[runningheads,a4paper]{llncs}
\usepackage{graphicx}
\usepackage{amsmath,amssymb}
\usepackage{xcolor}
\usepackage{booktabs}
\usepackage{tikz}
\usepackage{verbatim}
\usepackage{algorithm}
\usepackage{hyperref}
\usepackage{listings}

\hypersetup{
colorlinks=true,
linkcolor=black,
citecolor=black,
filecolor=blue,
urlcolor=blue
}

\newcommand{\class}[1]{{\bf #1}}

\newtheorem{defi}{Definition}
\newtheorem{thm}{Theorem}
\newtheorem{lem}[thm]{Lemma}

\newtheorem{cor}[thm]{Corollary}

\newtheorem{ex}{Example}

\newcommand{\ignore}[1]{}

\newcommand{\QED}{\hspace*{\fill}$\Box$}
\newcommand{\QEX}{\hspace*{\fill}$\blacksquare$}

\newenvironment{enumerate-} % low-profile enumerate
{\begin{enumerate}
    
   \setlength{\parskip}{-1ex}              % vertical space between paragraphs
   \setlength{\itemsep}{1.5ex}             % vertical space between items
}
{
 \end{enumerate}
}

\newcommand{\K}{{\mathcal K}}
\newcommand{\T}{{\mathcal T}}
\newcommand{\pK}{\Psi_{\K}}

\newcommand{\alg}{\mathcal} %'algebra' of given carrier set

\newcommand{\A}{\alg{A}}
\newcommand{\B}{\alg{B}}
\newcommand{\Pred}{{\sf Pred}}

\begin{document}
\title{Symbol Elimination for Parametric Second-Order Entailment Problems\\
{\small (with Applications to Problems in Wireless Network Theory)}}
\titlerunning{Symbol Elimination and Applications}
\author{Dennis Peuter, Philipp Marohn and Viorica
 Sofronie-Stokkermans
 %\orcidID{0000-0002-8486-9955}  
 %\and Dennis Peuter\inst{1}
 }
% \href{https://orcid.org/0000-0002-8486-9955}{\includegraphics[scale=1]{orcid.png}}

% \author{~~
% First Author\inst{1}\orcidID{0000-1111-2222-3333} \and
% $ Third Author\inst{3}\orcidID{2222--3333-4444-5555}
% }

\institute{ University Koblenz-Landau, Germany
% ~~~ \\
}

\maketitle              % typeset the header of the contribution

\begin{abstract}
We analyze possibilities of second-order quantifier elimination for
formulae containing parameters --
constants or functions. For this, we use a constraint
resolution calculus obtained from specializing the hierarchical
superposition calculus. 
If saturation terminates, we analyze possibilities of obtaining
weakest constraints on parameters which guarantee satisfiability. 
If the saturation does not terminate, we identify situations in which 
finite representations of infinite saturated sets exist.
We identify situations in which entailment between 
formulae expressed using second-order quantification can be
effectively checked.
We illustrate the ideas on a series of examples from wireless network research.
\end{abstract}

\section{Introduction}

The main motivation for this work was a study of models for graph classes naturally occurring in 
wireless network research -- in which nodes that are close are always connected, nodes that 
are far apart from each other are never connected and any other node pairs can, but do not need to 
be connected. 
Transformations can be applied to such graphs 
to make them symmetric;  
this way we can define further graph classes.   
When checking inclusion between graph
classes described using transformations 
we need to check entailment of second-order formulae.
In addition, many such graph class descriptions are parametric in
nature, so the goal is, in fact, to obtain (weakest) conditions on the 
parameters used in such descriptions that guarantee that graph classes 
are non-empty or that inclusions hold. This can be achieved by eliminating 
``non-parametric'' constants or function symbols used in the
description of such classes.

\smallskip
\noindent In this paper we combine methods for general symbol
elimination (which we use for eliminating existentially quantified
predicates) with methods for property-directed symbol elimination 
(which we use for obtaining conditions on ``parameters'' under which 
formulae are satisfiable or second-order entailment
holds). For general second-order quantifier elimination we use 
a form of ordered resolution similar to that proposed in \cite{Gabbay-Ohlbach}. 
For property-directed symbol elimination we use a method we proposed
in \cite{Sofronie-lmcs2018}. The advantage of using such a two-layered
approach is that it avoids non-termination that might occur if using only
general symbol elimination methods. The main application area 
we consider in this paper is the analysis of inclusions between graph
classes arising in wireless network research. 
Our main contributions are: 
\begin{itemize}
\item We analyze theories used in modeling graph classes and 
prove locality of theories of ``distances'' occurring in this
context. 
\item We analyze possibilities of general symbol elimination, using a simple 
specialization $\mathit{HRes}^P_{\succ}$ of the hierarchical
 superposition calculus 
(a form of ordered resolution) for eliminating a predicate symbol $P$. 
\item If saturation terminates, we analyze
  possibilities of obtaining weakest constraints on parameters
  occurring in the clauses which guarantee satisfiability, using
  methods for property-directed symbol elimination. 
\item If the saturation does not terminate, we study possibilities
of representing an (infinite) saturated set as a set of constrained
clauses in which the constraints are
interpreted in the minimal model of a set of constrained Horn
clauses. 
\item We analyze possibilities of effectively
checking entailment between formulae expressed using second-order 
quantification. 
\item We illustrate the ideas on examples related to the study
  of wireless networks. 
\end{itemize}

\

\noindent {\em Related work.} 
The study of second-order quantifier elimination 
goes back to the beginning of the 20th century (cf.\ \cite{Behmann22,Ackermann35a,Ackermann35b}).
Most of its known applications 
are in the study of modal logics or knowledge representation 
\cite{GabbaySchmidtSalasz,GorankoHustadtSchmidtVakarelov}; 
in many cases second-order quantifier elimination is proved only 
for very restricted fragments (cf.\ e.g.\ \cite{Voigt}). 
In \cite{Gabbay-Ohlbach}, Gabbay and Ohlbach proposed a
resolution-based algorithm for second-order quantifier elimination
which is implemented in the system SCAN. 
In \cite{BGW94}, Bachmair et al.\ mention that 
hierarchical superposition (cf.\ 
\cite{BaumgartnerWaldmann13,BaumgartnerWaldmann2019} for further refinements) 
can be used for second-order quantifier elimination modulo a theory. 
In \cite{Voronkov09,Voronkov10}, Hoder et al. study 
possibilities of symbol elimination in inference systems (e.g.\ the superposition calculus and its extension with 
ground linear rational arithmetic and uninterpreted functions). 
The main challenge when using saturation approaches for symbol
elimination is the fact that the saturated sets might be infinite. 
Sometimes finite representations of possibly infinite sets of clauses exist:
for this, Horbach and Weidenbach introduced a melting calculus
\cite{Horbach}, later used in 
\cite{Horbach-Sofronie13,Horbach-Sofronie14} and \cite{FietzkeWeidenbach}.  
Similar aspects were explored in the study of {\em acceleration} 
for program verification modulo Presburger arithmetic by 
Boigelot,  Finkel and Leroux \cite{Boigelot,FinkelLeroux}, in relationship  with array
systems by \cite{GhilardiSharyghina}, 
or in the study of constrained Horn clauses 
(cf. e.g. the survey \cite{McMillanRybalchenkoBjorner}).

\medskip
\noindent Orthogonal to this direction of study is what we call 
``property-directed'' symbol elimination: There, given
a theory $\T$ and a ground formula $G$ satisfiable
w.r.t.\ $\T$, the goal is to derive a (weakest) universal formula
$\Gamma$ 
over a  subset of the signature, such that $\Gamma \wedge G$ is
unsatisfiable w.r.t.\ $\T$. We devised methods for solving such
problems in \cite{Sofronie-lmcs2018} and used them for interpolant
computation \cite{Sofronie-lmcs2018}, and invariant
generation \cite{PeuterSofronie2019}.

\medskip
\noindent We are not aware of other similar approaches to the area of computational (geometric) 
graph theory. Existing approaches use a logical representation of graphs based on 
monadic second-order logic (cf.\ e.g.\ \cite{Courcelle97}) or 
higher-order theorem provers like Isabelle/HOL (cf.\ e.g.\
\cite{MehlhornNipkow19}). 
Our approach is orthogonal; it allows a reduction of many 
problems to satisfiability modulo a suitable theory.

\medskip
\noindent {\em Structure of the paper.} 
In Section~\ref{networks} we present the motivation for our research.
In Section~\ref{prelim} we introduce the notions on (local) theory
extensions needed in
the paper and prove the locality of theories of distance functions. 
In Section~\ref{prop-dir-symb-elim} we describe (and slightly extend)
a method for property-directed symbol elimination we proposed in \cite{Sofronie-lmcs2018}.
In Section~\ref{second-order-qe} we present the
$\mathit{HRes}^P_{\succ}$ calculus we use for eliminating predicate $P$, and 
analyze possibilities of giving finite representations 
for infinite saturated sets and of investigating the satisfiability of the saturated
sets. In Section~\ref{class-containment} we use these
ideas for checking class inclusion. In Section~\ref{tests} we discuss 
the way in which we tested the methods we propose on various examples.
In Section \ref{conclusions} we
present conclusions and plans for future work.

\medskip
\noindent 
This paper is an extended version
of~\cite{peuter-sofronie-frocos-2021} which contains full proofs of
the results, a more detailed description of the examples, a
description of the systems we used for testing and several examples
illustrating how these systems were used. 

\

\noindent {\bf Table of Contents}

\medskip

\contentsline {section}{\numberline {1}Introduction}{1}{section.1.1}
\contentsline {section}{\numberline {2}Motivation}{3}{section.1.2}
\contentsline {section}{\numberline {3}Theories and local theory extensions}{5}{section.1.3}
\contentsline {subsection}{\numberline {3.1}Local theory extensions}{6}{subsection.1.3.1}
\contentsline {subsection}{\numberline {3.2}Locality of theories of distances}{10}{subsection.1.3.2}
\contentsline {section}{\numberline {4}Property-directed symbol elimination and locality}{14}{section.1.4}
\contentsline {section}{\numberline {5}Second-order quantifier elimination}{15}{section.1.5}
\contentsline {subsection}{\numberline {5.1}Case 1: Saturation is finite}{19}{subsection.1.5.1}
\contentsline {subsection}{\numberline {5.2}Case 2: Finite representation of possibly infinite saturated sets}{23}{subsection.1.5.2}
\contentsline {section}{\numberline {6}Checking  Entailment}{26}{section.1.6}
\contentsline {subsection}{\numberline {6.1}Application: Checking class inclusion}{28}{section.1.6.1}
\contentsline {section}{\numberline {7}Tests}{29}{section.1.7}
\contentsline {section}{\numberline {8}Conclusions}{30}{section.1.8}
\contentsline {section}{\numberline {A}Tests}{35}{section.A.1}
\contentsline {subsection}{\numberline {A.1}Tests for Example \ref {ex-sat2}}{35}{subsection.A.1.1}
\contentsline {subsection}{\numberline {A.2}Tests for Example \ref {ex:sat}}{36}{subsection.A.1.2}
\contentsline {subsection}{\numberline {A.3}Tests for Example \ref {qudg}}{39}{subsection.A.1.3}
\contentsline {section}{\numberline {B}Proof of Theorem\nobreakspace {}\ref {non-linear}}{43}{section.A.2}
\contentsline {section}{\numberline {C}Constrained Horn clauses: Definitions}{44}{section.A.3}

\section{Motivation} 
\label{networks}

\noindent {\bf Graph Classes.} 
Graph classes important in wireless network research 
are: The class $\class{UDG}$ of  \emph{unit disk graphs}
(two nodes are connected iff they are different and their distance is $\leq 1$); 
the class $\class{QUDG}(r)$ of quasi unit disk graphs, for $r
  \in (0, 1]$ (two distinct nodes with distance  $\leq r$ 
are always connected and nodes with distance $> 1$ 
are never connected); 
the class $\class{DTG}(\mathbf{r}, r)$ of {\em directed transmission
    graphs} for $r > 0$ (every node $v$ has
a maximum communication distance ${\mathbf r}(v) \leq r$;
an edge from $v$ to $w$ exists iff $v \neq w$ and the distance
between $v$ and $w$ is $\leq {\mathbf r}(v)$). 

\smallskip
\noindent Many graph classes $\class{C}({\overline p})$ (where
${\overline p}$ is a sequence of symbols denoting {\em parameters}) 
can be described  using {\em inclusion}, {\em  exclusion} and 
{\em transfer} axioms. 

\noindent The {\em inclusion axioms} specify which edges have to
exist. For a graph class $\class{C}$ the condition under which an 
edge $E(u,v)$ must exist can be described by a formula $\pi_C^i(u,v)$. 
Therefore, inclusion axioms have the form: 

\medskip
$(1) \quad \quad \forall u, v ~(\pi_C^i(u,v) \rightarrow E(u,v))$ 

\medskip
\noindent The {\em exclusion axioms} specify which edges are not allowed to exist. 
For a class $\class{C}$ the condition under which an edge $E(u,v)$ is
not allowed  to exist  can be described by a formula $\pi_C^e(u,v)$. 
Therefore, transfer axioms have the form: 

\medskip
$(2) \quad \quad  \forall u, v ~(\pi_C^e(u,v) \rightarrow \lnot E(u,v))$ 

\medskip
\noindent The {\em transfer axioms} specify which edges $E(u,w)$ must 
exist as a consequence of the existence of another edge $E(u,v)$. 
For a class $\class{C}$, we describe these these 
conditions by a formula $\pi_C^t(u,v,w)$.
Therefore, inclusion axioms have the form: 

\medskip
$(3) \quad \quad \forall u, v, w ~ \pi_C^t(u,w,v) \land E(u,v)  \rightarrow E(u,w)$.

\

\noindent If the description of the graph class $\class{C}$ depends on
parameters ${\overline p}$, the formulae $\pi_C^i, \pi_C^e$ and
$\pi_C^t$ might contain parameters. We will sometimes indicate this by 
adding the parameters to the arguments, i.e.\ writing 
$\pi_C^i(u, v, {\overline p}), \pi_C^e(u, v, {\overline p})$ resp.\ 
$\pi_C^t(u, v, w, {\overline p})$.

\medskip 
\noindent We can, e.g., define the classes $\class{MinDG}(r)$, 
$\class{MaxDG}(r)$ and $\class{CRG}$ using axioms: 
\begin{itemize}
\item ${\sf MinDG}(r)$: axiom $(1)$, where $\pi^i(u, v, r)$ is the
  formula 
$u \neq v \land d(u,v) \leq r$; 
\item ${\sf MaxDG}(r)$: axiom $(2)$, where $\pi^e(u, v, r)$ is the
  formula 
$d(u,v) > r$; 
\item ${\sf CRG}$: axiom $(3)$, where
$\pi^t(u,w,v)$ is the formula $ u \neq w \land d(u,w) \leq d(u,v)$,
\end{itemize}
(where $r$ is supposed to be a parameter).

\smallskip 
\noindent With this notation, the inclusion axiom 
${\sf MinDG}(r)$ states that if $u \neq v$
and $d(u,v) \leq r$ an edge from $u$ to $v$ must exist; 
the exclusion axiom ${\sf MaxDG}(r)$ states that if $d(u,v) > r$ then we are not 
allowed to have an edge from $u$ to $v$. 
The transfer axiom ${\sf CRG}$
states that if $u$ and $w$ are different and there is an edge from 
$u$ to $v$ and $d(u,w) \leq d(u,v)$ then there must exist an edge 
also from $u$ to $w$.

\medskip
\noindent By combining such axioms we obtain axiomatizations for new
graph classes. If the classes $\class{A}$ and $\class{B}$ of graphs are axiomatized 
by axioms ${\sf Ax_A}$ and ${\sf Ax_B}$ then ${\sf Ax_A} \wedge {\sf
  Ax_B}$ is an axiomatization for the intersection $\class{A} \cap
\class{B}$. \\
For instance, the class 
$\class{UDG} = \class{MinDG}(1) \cap \class{MaxDG}(1)$
is axiomatized by ${\sf MinDG}(1) \wedge {\sf MaxDG}(1)$. 

\medskip
\noindent We may want to check whether a graph class $\class{C}({\overline p})$ 
has non-empty models, or to determine (weakest) conditions on 
the parameters ${\overline p}$ 
under which this is the case. This is one of the problems which will
be analyzed in this paper. 

\

\noindent {\bf Simple transformations on graph classes.} 
We can define transformations $\gamma$ on graphs that transform the
edges and leave the set of vertices unchanged, and form graph classes 
$$\gamma(\class{C}) = \{ \gamma(G) \mid G \in \class{C} \}.$$ 
Two examples of transformations are $\cdot^+$ and $\cdot^-$: 
Given a graph $G=(V,E)$, we can build the symmetric
supergraph $G^+=(V,E^+)$ resp.\ symmetric subgraph $G^-=(V,E^-)$,
defined by: 
$$ \begin{array}{l} 
\forall x, y ~(E^+(x, y)  \leftrightarrow (E(x, y) \lor E(y, x))) \\
\forall x, y ~(E^-(x, y)  \leftrightarrow  (E(x, y) \land E(y, x))).
\end{array}$$
We can thus define the classes $\class{C}^+ = \{ G^+ \mid G \in
\class{C} \}$ and $\class{C}^- = \{ G^- \mid G \in \class{C} \}$. 

\smallskip
\noindent The class of quasi unit disk graphs \cite{Barriere2003,Kuhn2008} can, 
for instance, be described as 
$$\class{QUDG}(r) = ( \class{MinDG}(r) \cap \class{MaxDG}(1) )^-.$$  
We might want to obtain an axiomatization for $\class{QUDG}(r)$ that
depends only on the predicates $\pi^i(x, y, r), \pi^e(x, y, 1)$ or test
whether the class is the same as the class described by
$(\class{MinDG}(r) \cap \class{MaxDG}(1) )^+$.

\smallskip
\noindent 
To find an axiomatization of a graph class $\gamma(\class{C})$, where $\gamma$ is a
transformation, we need to find a first-order formula equivalent to 
$\exists E' ( {\sf N}_{E'} \cap {\sf Tr}(E', E) ),$  where ${\sf
  N}_{E'}$ is a class of  clauses describing class $\class{C}$
and ${\sf Tr}$ is a formula describing the way the edges of the graph $(V,
E) = \gamma(V, E')$ can be obtained from the description of the graph
$(V, E')$. We here analyze possibilities of eliminating second-order quantifiers.

\medskip
\noindent {\bf Checking class inclusion.} If we can find such formulae for two graph
classes, then we can also check containment (provided the formulae
belong to decidable theory fragments). In this paper we analyze
situations in which this is possible.

\section{Theories and local theory extensions}
\label{prelim}

We assume known the basic notions in (many-sorted) first-order logic.
We consider signatures of the form $\Pi = (S, \Sigma, {\sf Pred})$, 
where $S$ is a set of sorts, $\Sigma$ is a family of function symbols and ${\sf Pred}$
a family of predicate symbols, such that for every function symbol $f$
(resp.\ predicate symbol $p$) their arity $a(f) = s_1
\dots s_n \rightarrow s$ 
(resp.\ $a(p) = s_1 \dots
s_m$), where $s_1, \dots, s_n, s \in S$,  is specified. 
If $C$ is a fixed countable set of fresh constants, we denote by 
$\Pi^C$ the extension of $\Pi$ with constants in $C$. 
We assume known standard definitions from first-order logic  
such as $\Pi$-structure, model,
satisfiability, unsatisfiability.
A $\Pi$-structure is a tuple 
$${\mathcal A} = (\{ A_s \}_{s \in S}, \{ f_{\mathcal A} \}_{f \in \Sigma}, \{
p_{\mathcal A} \}_{p \in {\sf Pred}}),$$ 
where, for every $s \in S$, $A_s$ is a non-empty set (the universe of sort $s$ of
the structure), for every $f \in \Sigma$ with arity $s_1 \dots s_n
\rightarrow s$, $f_{\mathcal A} : A_{s_1} \times \dots \times A_{s_n}
\rightarrow A_s$, 
and for every $p \in {\sf Pred}$ with arity $s_1 \dots s_m$, 
$p_{\mathcal A} \subseteq A_{s_1} \times \dots \times A_{s_m}$. 

\smallskip
\noindent If $\A$ is a $\Pi$-structure, we will denote by $\A^A$ the extension of
$\A$, where we have an additional constant (of sort $s$) for each element 
$a$ of sort $s$ of $\A$ (which we denote with the same symbol) with the 
natural interpretation
mapping the constant $a$ to the element $a$ of $\A$.

\smallskip
\noindent If $\Pi \subseteq \Pi'$ and $\A$ is
a $\Pi'$-structure, we denote its reduct to $\Pi$ by $\A_{|\Pi}$.

\medskip
\noindent {\bf Notation.} We will denote with (indexed
versions of) $x, y, z$ variables and with (indexed versions of) $a, b,
c, d$ constants; ${\overline x}$ will stand for a sequence of variables $x_1, \dots, x_n$, 
and ${\overline c}$ for a sequence of constants $c_1, \dots, c_n$. 

\medskip
\noindent {\bf Theories.} Theories can be defined by specifying a set
of axioms, or by specifying a
class of structures (the models of the theory). 
If $F$ and $G$ are formulae we 
write $F \models G$ (resp. $F \models_{\cal T} G$ -- also written as ${\cal T} \cup F \models G$) 
to express the fact that every model of $F$ (resp. every model of $F$ which is also a model of 
$\T$) is a model of $G$.  We denote
``falsum'' with $\perp$. $F \models \perp$ means that $F$ is
unsatisfiable; $F \models_{\T} \perp$ means that there is no model of
$\T$ in which $F$ is true. 

\smallskip
\noindent 
A theory $\T$ over a signature  
$\Pi$ {\em allows quantifier elimination} (QE) if for every formula $\phi$ over  
$\Pi$ there exists a quantifier-free formula $\phi^*$ over  
$\Pi$ which is equivalent to $\phi$ modulo $\T$. 
% \end{defi} 
Examples of theories which allow quantifier elimination are
rational and real linear arithmetic (${\sf LI}({\mathbb Q})$, ${\sf
  LI}({\mathbb R})$), 
the theory of real closed fields, 
and the theory of absolutely-free data structures.

\medskip
\noindent 
Sometimes, in order to define
more complex theories we can consider theory extensions and
combinations thereof. {\em Local theory extensions} are a class of 
theory extensions for which hierarchical reasoning is possible.

\subsection{Local theory extensions}  
In what follows, for simplicity we present the main notions in the
one-sorted case; the extension to the many-sorted case is immediate.

\smallskip
\noindent 
Let $\Pi_0 {=} (\Sigma_0, {\sf Pred})$ be a signature, and ${\mathcal T}_0$ be a 
``base'' theory with signature $\Pi_0$. 
We consider 
extensions $\T := {\mathcal T}_0 \cup \K$
of ${\mathcal T}_0$ with new function symbols $\Sigma$
({\em extension functions}) whose properties are axiomatized using 
a set $\K$ of (universally closed) clauses 
in the extended signature $\Pi = (\Sigma_0 \cup \Sigma, {\sf Pred})$, 
such that each clause in $\K$ contains function symbols in $\Sigma$. 
Especially well-behaved are the {\em $\Psi$-local theory extensions}, i.e.\
theory extensions $\T_0 \subseteq \T_0 \cup \K$ as defined above, 
in which checking ground satisfiability can be done using a finite
instantiation 
scheme described by a suitable closure operator $\Psi$, without loss
of completeness. We express this with the following condition:

\begin{tabbing}
\= ${\sf (Loc}^\Psi_f)$~ \quad \quad  \= For every finite set $G$ of ground
$\Pi^C$-clauses (for an additional \\
\> \> set $C$ of constants) it holds that $\T_0 \cup {\mathcal K} \cup
G \models \bot$  if and only if \\
\> \> $\T_0
\cup \K[\Psi_\K(G)] \cup G$ is unsatisfiable. 
\end{tabbing}

\noindent 
where, for every set $G$ of ground $\Pi^C$-clauses,  
$\K[\Psi_\K(G)]$ is the set of instances of $\K$ in which the terms 
starting with a function symbol in $\Sigma$ are in $\Psi_{\K}(G) = \Psi({\sf
  est}(\K, G))$, 
where ${\sf est}(\K, G)$ is the set of ground terms starting with a 
function in $\Sigma$ occurring in $G$ or $\K$.

If $\T_0$ is the pure theory of equality, we obtain the notion of
locality \cite{McAllester-acm-tocl-02,Ganzinger-01-lics}. 

\medskip
\noindent 
{\bf Partial and total models.} 
In \cite{sofronie-cade-05} we showed that 
local theory extensions can be 
recognized by showing that certain partial models embed into total
ones, and in \cite{ihlemann-sofronie-2010} we established similar
results for $\Psi$-local theory extensions and generalizations
thereof. 
We introduce the main  
definitions here, following mainly the presentation from
\cite{ihlemann-sofronie-2010} and \cite{sofronie-lmcs-2018}.

Let $\Pi = (\Sigma, {\sf Pred})$ be a
first-order signature with set of function symbols $\Sigma$ 
and set of predicate symbols ${\sf Pred}$. 
A \emph{partial $\Pi$-structure} is a structure 
$\A = (A, \{f_\A\}_{f\in\Sigma}, \{p_\A\}_{p\in \Pred})$, 
where $A$ is a non-empty set, for every $n$-ary $f \in \Sigma$, 
$f_\A$ is a partial function from $A^n$ to $A$, and for every $n$-ary 
$p \in {\sf Pred}$, $p_\A \subseteq A^n$. We consider constants (0-ary functions) to be always 
defined. $\A$ is called a \emph{total structure} if the 
functions $f_\A$ are all total. 
Given a (total or partial) $\Pi$-structure $\A$ and $\Pi_0 \subseteq \Pi$ 
we denote the reduct of 
$\A$ to $\Pi_0$ by $\A{|_{\Pi_0}}$.

The notion of evaluating a term $t$ with variables $X$ w.r.t.\ 
an assignment $\beta : X \rightarrow A$
 for its variables in a partial structure
$\A$ is the same as for total algebras, except that the evaluation is
undefined if $t = f(t_1,\ldots,t_n)$ 
and at least one of
$\beta(t_i)$ is undefined, or else $(\beta(t_1),\ldots,\beta(t_n))$ is
not in the domain of $f_\A$.

\begin{defi}
A \emph{weak $\Pi$-embedding} between two partial $\Pi$-structures
$\A$ and $\B$, where $\A = (A, \{f_\A\}_{f\in \Sigma}, \{p_\A\}_{p \in \Pred})$
 and  $\B = (B, \{f_\B\}_{f\in \Sigma}, \{p_\B\}_{p \in \Pred})$
is a total map $\varphi : A \rightarrow B$ such that 
\begin{enumerate}
\item[(i)] $\varphi$ is
an embedding w.r.t.\ ${\sf Pred} \cup \{ = \}$, i.e.\ 
 for every $p \in {\sf Pred}$ with arity $n$ and every 
$a_1,\dots,a_n \in \A$, 
$(a_1,\dots,a_n) \,{\in}\, p_\A$ if and only if $(\varphi(a_1), \dots,
\varphi(a_n))\,{\in}\, p_\B$. 
\item[(ii)]  
whenever $f_\A(a_1, \dots, a_n)$ is defined (in $\A$),  then 
$f_\B(\varphi(a_1), \dots, \varphi(a_n))$ is defined (in $\B$) and 
$\varphi(f_\A(a_1, \dots, a_n)) = f_\B(\varphi(a_1), \dots, \varphi(a_n))$,
for all $f \in \Sigma$. 
\end{enumerate}
\end{defi}

\begin{defi}[Weak validity]
Let $\A$ be a partial $\Pi$-algebra
and $\beta : X {\rightarrow} A$ a valuation for its variables.
$(\A, \beta)$ {\em weakly satisfies a clause  $C$} (notation: 
$(\A, \beta) \models_w C$) if either some of the literals in 
$\beta(C)$ are not defined or otherwise all literals are defined and
for at least one literal $L$ in $C$, $L$ is true in $\A$ w.r.t.\ $\beta$. 
$\A$ is a {\em weak partial model} 
of a set of clauses ${\mathcal K}$ if $(\A, \beta)  \models_w C$ for every 
valuation 
$\beta$ and every clause $C$ in ${\mathcal K}$. 
\end{defi}

\medskip
\noindent {\bf Recognizing $\Psi$-local theory extensions.} 
In \cite{sofronie-cade-05} we proved that if 
every weak partial model of an extension 
${\mathcal T}_0 \cup {\mathcal K}$ of a base theory ${\mathcal T}_0$ 
with total base functions can be embedded into a total
model of the extension, then the extension is local. 
In \cite{sofronie-tacas08} we lifted these results to $\Psi$-locality. 

Let $\alg{A} = (A, \{ f_{\A} \}_{f \in \Sigma_0 \cup \Sigma} \cup C, \{ p_\A
\}_{p \in {\sf Pred}})$ be a partial $\Pi^C$-structure with total
$\Sigma_0$-functions. 
Let $\Pi^A$ be the extension of the signature $\Pi$ with constants 
from $A$. We denote by $T(\A)$ the following set of ground 
$\Pi^A$-terms: 
$$T(\A) := \{ f(a_1,...,a_n) \,|\; f \in \Sigma, a_i \in A, i=1,\dots,n, f_{\A}(a_1,...,a_n) \text{ is defined }  \}. $$
Let ${\sf PMod}_{w,f}^\Psi({\Sigma}, {\mathcal T})$ be the class of all
weak partial models $\A$ of ${\mathcal T}_0 \cup {\mathcal K}$, such that
$\A{|_{\Pi_0}}$ is a total model of $\T_0$, the
$\Sigma$-functions are possibly partial, $T(\A)$ is finite and 
all terms in $\Psi({\sf est}(\K, T(\A)))$ are
defined (in the extension $\A^A$ with
constants from $A$).
We consider the following embeddability property of partial algebras:

\medskip
\begin{tabbing}
\= $({\sf Emb}_{w,f}^\Psi)$ \quad \= Every $\alg{A} \in {\sf PMod}_{w,f}^\Psi({\Sigma},  \T)$ weakly embeds into a total model of $\T$.
\index{$\Psi$-!embeddability}
\end{tabbing}

\medskip
\noindent 
We also consider the property $({\sf EEmb}_{w,f}^{\Psi})$,  
which additionally requires the embedding to be {\em elementary}, and the property
$({\sf Comp}^{\Psi}_f)$, which requires that every  structure $\alg{A} \in {\sf
  PMod}_{w,f}^\Psi({\Sigma}, \T)$ embeds
into a total model of $\T$ {\em with the same support}. 
If $\Psi$ is the identity, we refer to these properties as
$({\sf Emb}_{w,f})$, $({\sf EEmb}_{w,f})$ and ${\sf Comp}_f$.

\

\noindent 
When establishing links between locality and embeddability we require 
that the clauses in $\K$
are \emph{flat} 
and \emph{linear} w.r.t.\ $\Sigma$-functions.
When defining these notions we distinguish between ground and 
non-ground clauses.

\begin{defi}
An {\em extension clause $D$ is flat} (resp. \emph{quasi-flat}) 
when all symbols 
below a $\Sigma$-function symbol in $D$ are variables (resp. variables or ground $\Pi_0$-terms).
$D$ is \emph{linear}  if whenever a variable occurs in two terms of $D$
starting with $\Sigma$-functions, the terms are equal, and 
no term contains two occurrences of a variable.

\noindent A {\em ground clause $D$ is flat} if all symbols below a $\Sigma$-function 
in $D$ are constants.
A {\em ground clause $D$ is linear} if whenever a constant occurs in two terms in $D$ whose root symbol is in $\Sigma$, the two terms are identical, and no term which starts with a $\Sigma$-function contains two occurrences of the same constant.
\label{flat}
\end{defi} 
 \begin{defi}[\cite{ihlemann-sofronie-2010}]
With the above notations, let $\Psi$ be a map associating with 
$\K$ and a set of $\Pi^C$-ground terms $T$ 
a set $\pK(T)$ of $\Pi^C$-ground terms. 
We call $\pK$ a \emph{term closure operator} if the following
holds for all sets of ground terms $T, T'$: 
\begin{enumerate}
\item $\mathrm{est}(\K, T) \subseteq \pK(T)$,
\item $T \subseteq T' \Rightarrow \pK(T) \subseteq \pK(T')$,
\item $\pK(\pK(T)) \subseteq \pK(T)$,
\item for
  any map $h: C \rightarrow C$, $\bar{h}(\pK(T)) = \Psi_{\bar{h}\K}(\bar{h}(T))$,
  where $\bar{h}$ is the canonical extension of $h$ to extension
  ground terms.
\end{enumerate}
\end{defi}

\begin{thm}[\cite{sofronie-tacas08,ihlemann-sofronie-2010}]
Let ${\mathcal T}_0$ be a first-order theory and $\K$ a set of universally closed flat clauses in the signature
$\Pi$. The following hold: 
\begin{enumerate}
\item If all clauses in $\K$ are linear  
and $\Psi$ is a term closure operator with the property 
that for every flat set of ground terms $T$, $\Psi(T)$ is
flat then either of the conditions $({\sf Emb}_{w,f}^\Psi)$ and $({\sf EEmb}_{w,f}^\Psi)$ implies
$({\sf Loc}_f^{\Psi})$.
\item If  the extension ${\mathcal T}_0 \subseteq {\mathcal T} {=} {\mathcal T}_0 {\cup} {\mathcal K}$
satisfies $({\sf Loc}_f^{\Psi})$ then $({\sf Emb}_{w,f}^\Psi)$ holds. 
\end{enumerate}
\label{check-loc}
\end{thm}
The linearity assumption needed to prove that
$({\sf Emb}_{w,f}^\Psi)$ implies $({\sf Loc}_f^{\Psi})$ can be relaxed
if the closure operator $\Psi$ has additional properties. 
\begin{thm}
Let ${\mathcal K}$ be a set of $\Sigma$-flat clauses, with the
property that every variable occurs only once in every term. 
Let 
$\Psi$ be a term closure operator with the property that 
for every flat set of ground terms $T$, $\Psi(T)$ is flat. 

\noindent Assume that $\K$ and $\Psi$
have the property that for every flat set of ground terms $T$ and 
for every clause $C \in \K$, 
if $C$ contains terms $f(x_1, \dots, x, \dots, x_n)$ and 
$g(y_1, \dots, x, \dots, y_m)$ (where $f, g \in \Sigma$ are extension
functions and $f$ and $g$ are not necessarily different), if 
$f(t_1, \dots, t, \dots, t_n),  g(s_1, \dots, s, \dots, s_m) \in \Psi_{\K}(T)$ then 
$f(t_1, \dots, s, \dots, t_n), g(s_1, \dots, t, \dots, s_m) \in \Psi_{\K}(T)$.
\noindent Then $({\sf Emb}_{w,f}^\Psi)$ implies $({\sf Loc}_f^{\Psi})$. 
\label{non-linear}
\end{thm}
{\em Proof:} The proof is included in 
Appendix~\ref{app:non-linear}.\footnote{A similar result can be proved also in the case in which 
some variables occur several times below a function symbol
if $\Psi_{\K}$ has the property that 
if $f(x_1, \dots, x, \dots, x, \dots x_n) \in \K$
and $f(t_1, \dots, s, \dots, t, \dots, t_n) \in \Psi_{\K}(T)$
then $f(t_1, \dots, t, \dots, t, \dots, t_n) \in \Psi_{\K}(T)$
and $f(t_1, \dots, s, \dots, s, \dots, t_n) \in \Psi_{\K}(T)$.} \QED 

\begin{thm}[\cite{sofronie-cade05,sofronie-tacas08}]
The following theory extensions have property $({\sf Comp}_f)$, hence are local: 
\begin{itemize}
\item[(i)] The extension of a theory $\T_0$ with uninterpreted
function symbols. 
\item[(ii)] The extension of a theory $\T_0$ containing a predicate
  $\leq$ which is reflexive with a function $f$ 
satisfying 
the axioms 
$\K  = \{
\forall {\overline x} ~ \phi_i({\overline x}) {\rightarrow}
L_i({\overline x}) \mid i = 1,\dots, n \},$ 
where:
\begin{itemize}
\item[--] $\phi_i$ are $\T_0$-formulae with $\phi_i({\overline x})
{\wedge} \phi_j({\overline x}) {\models} {\perp}$ if $i {\neq} j$\\
($n$ can be 1 and $\phi_1$ can be $\top$), 
\item[--] $L_i({\overline x})$ has the form 
~(1) $s_i \leq f({\overline  x})$ or 
~(2) $f({\overline x}) \leq t_i$ or 
~(3) $s_i \leq f({\overline  x}) \leq t_i$,  
~ where $s_i, t_i$ are $\Pi_0$-terms and  in case (3) $\phi_i \models_{\T_0} s_i
\leq t_i$.
\end{itemize}
\end{itemize}
\label{free-etc-local}
\end{thm}

\medskip
\noindent {\bf Hierarchical reasoning.}
Consider a $\Psi$-local theory extension 
${\mathcal T}_0 \subseteq {\mathcal T}_0 \cup {\mathcal K}$.
Condition $({\sf Loc}_f^{\Psi})$ requires that for every finite set $G$ of ground 
$\Pi^C$-clauses, ${\mathcal T}_0 \cup {\cal K} \cup G \models \perp$ iff 
${\mathcal T}_0 \cup {\mathcal K}[\Psi_{\cal K}(G)] \cup G \models \perp$.
In all clauses in ${\mathcal K}[\Psi_{\cal K}(G)] \cup G$ the function 
symbols in $\Sigma$ only have ground terms as arguments, so  
${\mathcal K}[\Psi_{\cal K}(G)] {\cup} G$ can be flattened 
and purified. 
We thus obtain a set of clauses ${\mathcal K}_0 \cup G_0 \cup {\sf Def}$, 
where ${\mathcal K}_0$ and $G_0$ do
not contain $\Sigma$-function symbols and ${\sf Def}$ contains clauses of the form 
$c = f(c_1, \dots, c_n)$, where $f \in \Sigma$, and $c, c_1, \dots,
c_n$ are constants.
This transformation allows us to reduce testing satisfiability w.r.t.\
${\mathcal T}_0 \cup {\mathcal K}$ to testing satisfiability w.r.t.\ ${\mathcal T}_0$.

\begin{thm}[\cite{sofronie-cade-05}]
Let ${\mathcal K}$ be a set of clauses. 
Assume that 
${\mathcal T}_0 \subseteq {\cal T}_1 = {\mathcal T}_0 \cup {\mathcal K}$ is a 
$\Psi$-local theory extension. 
For any finite set $G$ of ground $\Pi^C$-clauses, 
let ${\mathcal K}_0 \cup G_0 \cup {\sf Def}$ 
be obtained from ${\mathcal K}[\Psi_{\cal K}(G)] \cup G$ by introducing, in a bottom-up manner, new  
constants $c_t \in C$ for subterms $t {=} f(c_1, \dots, c_n)$ where $f {\in}
\Sigma$ and $c_i$ are constants, together with 
definitions $c_t {=} f(c_1, \dots, c_n)$ (included in ${\sf Def}$) and
replacing the corresponding terms $t$ with the constants $c_t$ in $\K$
and $G$. 
Then ${\cal T}_1 \cup G \models \perp$ if and only if 
${\mathcal T}_0 \cup {\mathcal K}_0 \cup G_0 \cup {\sf Con}_0 \models \perp,$ where 
$\displaystyle{{\footnotesize {\sf Con}_0  {=} \{ \bigwedge_{i = 1}^n c_i
    {\approx} d_i \rightarrow c {\approx} d \, {\mid}
\begin{array}{l}
f(c_1, \dots, c_n) {\approx} c {\in} {\sf Def}\\
f(d_1, \dots, d_n) {\approx} d {\in} {\sf Def} 
\end{array} \}}}.$\\[-3ex]
\label{lemma-rel-transl}
\end{thm} 
This method is implemented in the program H-PILoT (Hierarchical
Proving by Instantiation in Local Theory Extensions) \cite{hpilot}.

\subsection{Locality of theories of distances}

\noindent The theories related to wireless networks used in
Section \ref{networks} refer to cost or distance functions. 
We prove that axiomatizations for such functions define local theory extensions. 
We first formalize the properties of metric spaces $(X, d)$, i.e.\ 
sets endowed with a distance function $d$ satisfying the usual axioms 
of a metric, and prove a locality property. We then consider variants that contain only 
some of these axioms.

\begin{thm}
Let $\T_0$ be the disjoint two-sorted combination of ${\cal E}$, the pure theory of equality
(no function symbols), sort ${\sf p}$, and   $LI({\mathbb R})$
(linear real arithmetic), sort ${\sf num}$. 
Let $\T^m_d$ be the extension of $\T_0$ with a  function $d$ with arity $a(d) = {\sf p}, {\sf p}
\rightarrow {\sf num}$ satisfying the following set ${\cal K}_d$ of axioms: 
$$\begin{array}{lrl} 
(d_1) & \forall x, y \, & d(x, y) \geq 0 \\
(d_2) & \forall x, y, z \, & d(x, y) \leq d(x, z) + d(z, y) \\ 
(d_3) & \forall x, y \, &  d(x, y) = d(y, x) \\
(d_4) & \forall x, y \, & x = y \rightarrow d(x, y) = 0 \\
(d_5) &  \forall x, y \, & d(x, y) = 0 \rightarrow x = y 
\end{array}$$
Let $\Psi_m$ be defined for every set $T$ of ground terms  by 
\[ \Psi_m(T) = \{ d(a, b) \mid a, b \text{ are constants of sort }
{\sf p} \text{ occurring in } T \}. \]
Then the following hold: 
\begin{itemize}
\item[(1)] $\Psi_m$ is a closure operator on ground terms. 
\item[(2)] For every finite set $T$ of ground terms, $\Psi_m(T)$ is
  finite. 
\item[(3)] $\T^m_d$ is a $\Psi_m$-local extension of $\T_0$ satisfying
  condition $({\sf Comp}^\Psi_f)$. 
\end{itemize}
\end{thm} 
{\em Proof:}
(1) Clearly, for every set $T$ of ground terms, $T$ and 
$\Psi_m(T)$ contain the same constants of sort ${\sf p}$, so 
$\Psi_m(\Psi_m(T)) = \Psi_m(T)$. Since the only extension function
symbol is $d$, ${\sf est}(\K, T) \subseteq \Psi_m(T)$ for every set
$T$ of ground terms. The fact that if 
$T_1 \subseteq T_2$ we have $\Psi_m(T_1) \subseteq \Psi_m(T_2)$ follows
from the definition. It is also easy to check that for every map 
$h : C \rightarrow C$, ${\overline h}(\Psi_m(T)) = \Psi_m({\overline
  h}(T))$, i.e.\ $\Psi_m$ is stable under renaming of constants. 

 \medskip
\noindent (2) If $T$ is finite, then it contains finitely many
constants (say $n$). $\Psi_{m}(T)$ has then $n^2$ elements. 

\medskip
\noindent (3) To prove that $\T^m_d$ is a $\Psi_m$-local extension of $T_0$, 
we prove  that it satisfies the embeddability condition 
(${\sf Comp^{\Psi}_{w,f}}$), i.e.\ that 
for every partial model ${\cal P}  = (P, {\mathbb R}, d_P)$ of 
$\T_d = \T_0 \cup \K_d$ with the properties: 
\begin{itemize}
\item[(i)] All function symbols in $\Sigma_0$ are everywhere defined; $d$ is partially defined.
\item[(ii)] The set $T({\cal P}) = \{ d(a_1, a_2) \mid \text a_i \in A_p,
  d_P(a_1, a_2) \text{ is defined} \}$ is finite, and closed under $\Psi_m$. 
\end{itemize}
$d_P$ can be extended to a total function on $P$ that satisfies the axioms
$\K_d$.

Let ${\cal P} = (P, {\mathbb R}, d_P)$ be a partial model of $\T^m_d = \T_0
\cup \K_m$ (where $P$ is the support of sort {\sf p}, ${\mathbb
  R}$ the support of sort {\sf num}, and $d_P$ a partial function from
$P \times P$ to ${\mathbb R}$) satisfying the conditions above. Then: 
\begin{itemize}
\item whenever $d_P(p_1, p_2)$ defined, $d_P(p_1, p_2) \geq
0$, and if $p_1 = p_2$, $d_P(p_1, p_2) = 0$; 
\item $d_P(p, p)  = 0$ whenever it is defined; 
\item if $d_P(p_1, p_2)$ and $d_P(p_2, p_1)$ are defined then
  $d_P(p_1, p_2) = d_P(p_2, p_1)$; and 
\item if $d_P(p_1, p_2), d_P(p_2, p_3), d_P(p_1, p_3)$
are defined then  $d_P(p_1, p_2) \leq d_P(p_2, p_3) + d_P(p_1, p_3)$. 
\end{itemize}

\noindent 
Let $E = \{ (p_1, p_2) \mid  d_P(p_1, p_2)
\text{ is defined} \}$. 

\noindent Let $P_1 = \{ p \in P \mid \exists q \in P: d_P(p, q) \text{ defined or
} d_P(q, p) \text{ defined } \}$.
By the assumption that $d_P(p_1, p_2)$ is defined only for finitely many tuples
$(p_1, p_2)$, $P_1$ is finite and by condition (ii) above (as $T({\cal
  P})$
is closed under $\Psi_m$), 
$E = P_1 \times P_1$. 
Thus, $P = P_1 \cup P_2$, such that 
$d_1 = {d_P}_{|P_1}$ is totally defined and $d$ is nowhere defined on
$P_{2}$ (for every two different elements $p_1, p_2 \in P_{2}$, $d_P(p_1, p_2)$ is
undefined and for every $p \in P_2$ there is no $q \in P_1$ such that 
$d_P(p, q)$ or $d_P(q, p)$ is defined). 

\smallskip
\noindent 
Since $P_1$ is finite, the maximum $m_1 = {\sf max} \{ d(p, q)
\mid p, q \in P_1 \}$ exists. 

\smallskip
\noindent Consider an arbitrary distance function $d_{2}$ on
$P_{2}$ such that ${\sf sup} \{ d_{2}(p_1, p_2) \mid p_1, p_2 \in
P_{2} \}$ is finite (such a function is guaranteed to exist, since the 
distance axioms are  consistent: We can for instance regard all points 
in $P_{2}$ as points in the unit circle and consider the euclidian distances between
these points). 
Thus, the distance function $d_{2}$ on $P_{2}$ is totally defined
and bounded. Let $m_2$ be such that $d_2(p, q) \leq m_2$ for all $p,
q \in P_2$.

\smallskip
\noindent We now show how to extend $d$ on $P_1 \cup P_2$. 
If $P_1$ or $P_2$ are empty we have a total extension of $d$ already. 
Assume they are both non-empty. Let $p_1 \in P_1$ and $p_2 \in P_2$. 
We construct a totally defined function $d : (P_1 \cup P_2)^2
\rightarrow {\mathbb R}$ as follows: 

\[ d(p, q) = \left\{ \begin{array}{ll}
d_1(p, q) & \text{ if } p, q \in P_1 \\
d_2(p, q) & \text{ if } p, q \in P_2 \\
d_0 + d_1(p, p_1) + d_2(p_2, q) & \text{ if } p \in P_1 \text{ and
}  q \in P_2 \\
d_0 + d_1(q, p_1) + d_2(p_2, p) &  \text{ if } p \in P_2 \text{ and
}  q \in P_1 
\end{array}
\right. \]

\noindent where $d_0 \in {\mathbb R}$, is such that $d_0 = {\sf m} + 1$, where 
${\sf m} = {\sf max}(m_1, m_2)$.

\noindent 
We show that $d$ is a total function that satisfies all the axioms $\K_d$: 
\begin{itemize}
\item It is clear that $d$ is a total function and that for all $x, y \in
P_1 \cup P_2, d(x, y) \geq 0$, i.e.\ it satisfies axiom $(d_1)$. 

\item Let $p \in P_1 \cup P_2$. Then $p \in P_i$, with $i = 1$ or $2$,  
and since $d_i$ satisfies axiom $(d_4)$,  $d(p, p) = d_i(p, p) = 0$. 
Thus $d$  satisfies axiom $(d_4)$ too.

\item Let $p, q \in P_1 \cup P_2$. If $p, q \in P_i$ for $i = 1$ or $2$, 
then $d(p, q) = d_i(p, q) = d_i(q, p) = d(q, p)$ -- since $d_i$
satisfies axiom $(d_3)$. 
If $p \in P_1, q \in P_2$ then $d(p, q) = d_0 + d_1(p, p_1) +
d_2(p_2, q) = d_0 + d_1(p_1, p) + d_2(q, p_2) = d(q, p)$; 
the case when $p \in P_2, q \in P_1$ is similar.
Thus $d$  satisfies axiom $(d_3)$ too.

\item Let $p, q \in P_1 \cup P_2$. If $p, q \in P_i$, with $i = 1$ or $2$, 
and $d(p, q) = 0$ then $d_i(p, q) = 0$, so as $d_i$ satisfies axiom 
$(d_5)$, $p = q$. 
If $p \in P_1$ and $q \in P_2$ or $p \in P_2$ and $q \in P_1$
then by definition $d(p, q) \geq d_0 > 0$, so we cannot have $d(p, q) = 0$.
Thus $d$  satisfies axiom $(d_5)$ too. 

\item We show that $d$ satisfies the triangle inequality (axiom
$(d_2)$). Let $p, q, r \in P_1 \cup P_2$. We show that $d(p,q) \leq d(p, r) + d(r, q)$. 
We distinguish the following cases: 
\begin{description}
\item[Case 1: $p \in P_1, q \in P_2$.] Then  $d(p, q) = d_0 + d_1(p,
  p_1) + d_2(p_2, q)$. 

\begin{description}
\item[Subcase 1.a: $r \in P_1$.] Then $d(p, r) + d(r, q) = d_1(p, r) +
d_0 + d_1(r, p_1) + d_2(p_2, q) \geq d_0 + d_1(p, p_1)  +
d_2(p_2, q) = d(p, q)$. 
\item[Subcase 1.b: $r \in P_2$.] Then $d(p, r) + d(r, q) = d_0 + d_1(p, p_1) +
d_2(p_2, r) + d_2(r, q) \geq d_0 + d_1(p, p_1)  +
d_2(p_2, q) = d(p, q)$. 
\end{description}

\item[Case 2: $p \in P_2, q \in P_1$.] Then  $d(p, q) = d_0 + d_1(q,
  p_1) + d_2(p_2, p)$. 

\begin{description}
\item[Subcase 2.a: $r \in P_2$.] Then $d(p, r) + d(r, q) = d_2(p, r) +
d_0 + d_1(r, p_2) + d_2(p_1, q) = d_2(r, p) +
d_0 + d_1(q, p_1) + d_2(p_2, r) \geq d_0 + d_1(q, p_1)  +
d_2(p_2, p) = d(p, q)$. 
\item[Subcase 2.b: $r \in P_1$.] Then $d(p, r) + d(r, q) = d_0 + d_2(p, p_2) +
d_2(p_1, r) + d_1(r, q)  \geq d_0 + d_1(p_1, q)  +
d_2(p, p_2)  = d_0 + d_1(q, p_1)  +
d_2(p_2, p) = d(p, q)$. 
\end{description}

\item[Case 3: $p, q \in P_1$.] Then  $d(p, q) = d_1(p,q)$.
 \begin{description}
\item[Subcase 3.a: $r \in P_1$.] Then $d(p, q) = d_1(p, q) \leq d_1(p,
  r) + d_1(r, q) = d(p, r) + d(r, q)$, since $d_1$ satisfies axiom $(d_2)$. 
\item[Subcase 3.b: $r \in P_2$.] Then $d(p, q) {=} d_1(p, q) {\leq} {\sf
    m} {<} d_0 {\leq} d(p,
  r) {+} d(r, q)$. 
\end{description}

\item[Case 4: $p, q \in P_2$.] Then  $d(p, q) = d_2(p,q)$.
 \begin{description}
\item[Subcase 4.a: $r \in P_2$.] Then $d(p, q) = d_2(p, q) \leq d_2(p,
  r) + d_2(r, q) = d(p, r) + d(r, q)$, since $d_2$ satisfies axiom $(d_2)$. 
\item[Subcase 4.b: $r \in P_1$.] Then $d(p, q) {=} d_2(p, q) {\leq} {\sf
    m} {<} d_0 {\leq} d(p,
  r) {+} d(r, q)$. 
\end{description}
\end{description}
\end{itemize}
In \cite{ihlemann-sofronie-2010} it was proved that 
condition $({\sf Comp}^{\Psi}_f)$ for $\T_0 \subseteq \T_0 \cup \K$ 
implies $\Psi$-locality of the extension if the clauses in
$\K$ are flat and linear. The clauses in $\K_m$ are flat, but are not
linear. In the proof of the fact that embeddability implies locality 
linearity is needed in order to ensure that if we have a model $\B$ of 
$\T_0 \cup \K[\Psi(G)] \cup G$ we can define a partial model $\A$ of 
$\T_0 \cup \K \cup G$ and argue that (by $({\sf Comp}^{\Psi}_f)$) this 
model embeds into a total model of $\T_0 \cup \K \cup G$. 
We construct $\A$ as follows: Its universe(s) are the same as for
$\B$, and $f(a_1, \dots, a_n)$ is defined in $\A$ if there exists 
constants $c_1, \dots, c_n$ which interpret in $\A$ as $a_1, \dots,
a_n$ and $f(c_1, \dots, c_n)$ occurs in $\Psi(G)$. 
This definition is used to associate with every valuation in $\A$ in
which all terms in a clause $C$ are defined a substitution $\sigma$ 
such that $C\sigma \in \K[\Psi(G)]$. 

If the clause $C$ is linear the substitution can be defined without
problems. If $C$ contains a variable in different terms, it might be 
difficult to define $\Sigma$ because for different occurrences of 
$x$ we might find different suitable terms. 

This problem does not occur here because of the fact that $\Psi_m$ adds 
all necessary instances that allow to define $\sigma$ without
problems. 

Alternatively, it can be easily checked that all 
assumptions in Theorem~\ref{non-linear} 
hold in this case, so in this case embeddability entails locality. 
\QED

\medskip

\noindent 
We can still obtain local theory extensions if we leave out some of
the metric axioms. Below we consider, for instance, extensions with a 
function $d$ in which all the axioms of a metric except for the 
triangle inequality hold. 
\begin{thm}
Let $\T_0$ be the disjoint two-sorted combination of the theory $\cal{E}$ of pure equality
(no function symbols), sort ${\sf p}$, and   $LI({\mathbb R})$
(linear real arithmetic), sort ${\sf num}$. 
Let $\T^n_d$ be the extension of $\T_0$ with a function $d$ with arity $a(d) = {\sf p}, {\sf p}
\rightarrow {\sf num}$ satisfying the following set ${\cal K}_n$ of axioms: 
$$\begin{array}{lrl} 
(d_1) & \forall x, y \, & d(x, y) \geq 0 \\
(d_3) & \forall x, y \, &  d(x, y) = d(y, x) \\
(d_4) & \forall x, y \, & x = y \rightarrow d(x, y) = 0 \\
(d_5) &  \forall x, y \, & d(x, y) = 0 \rightarrow x = y 
\end{array}$$
Let $\Psi_n$ be defined for every set $T$ of ground terms by 
\[ 
\begin{array}{ll}
\Psi_n(T) = T & \cup \{ d(t_2, t_1) \mid d(t_1, t_2) \in T \} \\
 & \cup \{ d(a, a) \mid a \text{ is a constant of sort {\sf p}
   occurring in } T \}
\end{array}\]
Then $\T^n_d$ is a $\Psi_n$-local extension of $\T_0$. 
\end{thm} 
{\em Proof:}
To prove locality we have to show that every partial model of $\T^n_d
= \T_0 \cap {\cal K}_n$ 
which is closed under $\Psi_n$ can be extended to a total model.
Let ${\cal P} = (P, {\mathbb R}, d_P)$ be a partial model of $\T_d = \T_0
\cup \K_d$ (where $P$ is the support of sort {\sf p}, ${\mathbb
  R}$ the support of sort {\sf num}, and $d_P$ a partial function from
$P \times P$ to ${\mathbb R}$) satisfying the conditions above. 
We construct a total function $d : P \times P
\rightarrow {\mathbb R}$ as follows: 
\[ d(p, q) = \left\{ \begin{array}{ll}
d_P(p, q) & \text{ if } d_P(p, q) \text{ is defined}  \\
0 & \text{ if } d_P(p, q) \text{ is not defined and } p = q \\
1 & \text{ if } d_P(p, q) \text{ is not defined and } p \neq q \\
\end{array}
\right. \]
It is easy to check that $d$ satisfies all the axioms in ${\cal
  K}_n$. 
The considerations in the previous proof (or Theorem~\ref{non-linear})
can be used also in this case
to show that embeddability implies locality in spite of the
non-linearity due to the choice of the closure operator. 
\QED

\begin{thm}
Let $\T_0$ be the disjoint combination  of the theory of pure
 equality (sort ${\sf p}$) and linear real arithmetic (sort ${\sf
 num}$).
The following extensions of $\T_0$ with a function $d$ (sort ${\sf p} {\times} {\sf p}
  {\rightarrow} {\sf num}$) are $\Psi$-local, with $\Psi$ being the
  identity function.  
\begin{itemize}
\item[(i)] $\T^u_d$, the extension of $\T_0$ with an uninterpreted
  function $d$. 
\item[(ii)] $\T^p_d = \T_0 \cup \K_p$, where $\K_p = \forall
  x,y ~ d(x, y) \geq 0$. 
\end{itemize}
The extension $\T^s_d = \T_0 \cup \K_s$, where $\K_s = \forall x, y ~
d(x, y) = d(y, x)$ is $\Psi_s$-local, where $\Psi_s(T) = T \cup \{
d(a, b) \mid d(b, a) \in T \}$. 
\label{loc-rest}
\end{thm}
{\em Proof.} (i) and (ii) are a direct consequence of
Theorem~\ref{free-etc-local}; the locality proof for $\T^s_d$ is
similar to the one for $\T^n_d$. \QED

\medskip
\noindent We present all the results together in the following theorem: 
\begin{thm}
Let $\T_0$ be the disjoint combination  of the theory ${\cal E}$ of pure
  equality (sort ${\sf p}$) and linear real arithmetic (sort ${\sf
    num}$). The following extensions of $\T_0$ with a function $d$ (sort ${\sf p} {\times} {\sf p}
  {\rightarrow} {\sf num}$) are $\Psi$-local for a suitable closure operator $\Psi$:
\begin{itemize}
\item[(1)] $\T^m_d = \T_0 \cup \K_m$, where $\K_m$ are axioms of a
  metric, is $\Psi_m$-local, where  
  $\Psi_m(T) = \{ d(a, b) \mid a, b \text{ constants of sort } {\sf p}
  \text{ occurring in } T \}$.
\item[(2)] $\T^n_d = \T_0 \cup \K_n$, where $\K_n$ contains 
  all axioms of a
  metric except for the triangle inequality, is $\Psi_n$-local, where  
  $\Psi_n(T) = T \cup \{ d(b, a) \mid d(a, b) \in T \} \cup \{ d(a, a) \mid
  a \text{ constant of sort } {\sf p} \text{ occurring in } T \}$.
\item[(3)] $\T^u_d$, the extension of $\T_0$ with an uninterpreted
  function $d$, and $\T^p_d = \T_0 \cup \K_p$, where $\K_p = \forall
  x,y ~ d(x, y) \geq 0$, are $\Psi$-local, where $\Psi(T) = T$. 
\end{itemize}
\label{thm:distance}
\end{thm}

\vspace{-7mm}

\section{Property-directed symbol elimination and locality}
\label{prop-dir-symb-elim}

\noindent In \cite{Sofronie-lmcs2018} we proposed a method for
property-directed symbol elimination described in Algorithm~1.
We present a slight generalization. 

\begin{algorithm}[t]
\caption{Symbol elimination in theory extensions
  \cite{Sofronie-ijcar16,Sofronie-lmcs2018}}
\label{algorithm-symb-elim} 
\begin{tabular}{ll} 
{\bf Input:} & Theory extension $\T_0 \cup \K$ with
signature $\Pi = \Pi_0 \cup (\Sigma \cup \Sigma_{\sf par})$ \\
                   & ~~where $\Sigma_{\sf par}$ is a set of parameters \\
                   & Set $T$ of ground $\Pi^C$-terms \\
{\bf Output:} & $\forall {\overline y} \Gamma_T({\overline y})$ (universal $\Pi_0 \cup \Sigma_{\sf par}$-formula)  
~~~~~~~~~~~~~~~~~~~~~~~~~~~~~~~~~~~~~~~~~~~~~~~~~\\
\hline 
\end{tabular}

{\small 
\begin{description}
\item[Step 1] Purify $\K[T] \cup G$ as described in Theorem~\ref{lemma-rel-transl} (with set of extension
  symbols $\Sigma_1$). 
Let $\K_0 \cup G_0 \cup {\sf Con}_0$ be the set of
  $\Pi_0^C$-clauses obtained this way. 

\smallskip
\item[Step 2] Let $G_1 = {\mathcal K}_0 \cup G_0\cup {\sf Con}_0$. 
Among the constants in $G_1$, we identify 
\begin{enumerate}
\item[(i)] the constants
$c_f$, $f \in \Sigma_{\sf par}$, where  $c_f$ is a constant
parameter or $c_f$ is 
introduced by a definition $c_f \approx f(c_1, \dots, c_k)$ in the hierarchical
reasoning method, % and 
\item[(ii)] all constants  ${\overline c_p}$ 
occurring as arguments of functions in $\Sigma_{\sf par}$ in such definitions. 
\end{enumerate}
Replace all the other constants ${\overline  c}$
with existentially quantified variables ${\overline x}$
(i.e.\ 
replace $G_1({\overline c_p}, {\overline c_f}, {\overline c})$ 
with $\exists {\overline x} G_1({\overline c_p},
{\overline c_f}, {\overline x})$).

\smallskip
\item[Step 3] Construct 
a formula  $\Gamma_1({\overline c_p}, {\overline c_f})$ equivalent to 
$\exists {\overline x} G_1({\overline c_p}, {\overline c_f},{\overline
  x})$
w.r.t.\ $\T_0$ using a method for quantifier elimination in 
${\mathcal T}_0$.

\smallskip
\item[Step 4] Replace 
each constant $c_f$ introduced by definition $c_f = f(c_1, \dots,
c_k)$ with the term $f(c_1, \dots,c_k)$ in $\Gamma_1({\overline c_p}, {\overline c_f})$.
Let $\Gamma_2({\overline c_p})$ be the formula 
obtained this way. 
Replace ${\overline c_p}$ with existentially quantified variables ${\overline y}$. 

\smallskip
\item[Step 5] Let $\forall {\overline y} \Gamma_T({\overline y})$ be
  $\forall {\overline y} \neg \Gamma_2({\overline y})$. 
\end{description}
} 
\end{algorithm}

\begin{thm}[\cite{Sofronie-ijcar16,Sofronie-lmcs2018}] 
Let ${\cal T}_0$ be a $\Pi_0$-theory allowing quantifier elimination\footnote{If $\T_0$ does not allow QE
  but has a model completion $\T_0^*$ which does, and 
  if we use QE in $\T_0^*$ in Algorithm 1, $\T_0 \wedge \forall
  {\overline x} \Gamma_T({\overline x}) \cup G \models \bot$, but $\forall
  {\overline x} \Gamma_T({\overline x})$ might not be the weakest
  universal formula $\Gamma$ with the property that $\T_0
  \cup \Gamma \cup \K \models \bot$.}
$\Sigma_{\sf par}$ be a set of parameters 
(function and constant symbols) and $\Pi = (S, \Sigma, {\sf Pred})$ 
be such that $\Sigma \cap (\Sigma_0 \cup \Sigma_{\sf par}) =
\emptyset$. 
Let ${\cal K}$ be a set of clauses 
in the signature $\Pi_0 {\cup} \Sigma_{\sf par} {\cup} \Sigma$ in which all 
variables occur also below functions in $\Sigma_1 = \Sigma_{\sf par} \cup
\Sigma$.  
Assume $\T \subseteq \T_0 \cup \K$ satisfies
condition $({\sf Comp}^{\Psi}_f)$ for a suitable closure operator
$\Psi$ with ${\sf est}(G) \subseteq \Psi_{\K}(G)$ for every set $G$ of
ground $\Pi^C$-clauses. Then, for $T = \Psi_{\K}(G)$, Algorithm~1 yields a universal 
$\Pi_0 \cup \Sigma_{\sf par}$-formula 
$\forall {\overline x} \Gamma_T({\overline x})$ 
such that ${\cal T}_0 \cup \forall {\overline x}
\Gamma_T({\overline x}) \cup {\cal K} \cup
G \models \perp$ which is entailed by every
universal formula $\Gamma$ with 
  ${\cal T}_0 \cup \Gamma \cup
  {\cal K} \cup G \models \perp$.
\label{thm-alg-1}
\end{thm}

\noindent {\em Proof:} 
The fact that $\T_0 \cup \K \cup \forall x \Gamma_T(x) \models \perp$ 
was proved in \cite{Sofronie-lmcs2018}. 
We show that if $T = \Psi_{\K}(G)$ then 
for every set $\Gamma$ of universal constraints on the parameters, 
if ${\mathcal T}_0 \cup \Gamma \cup {\mathcal K} \cup G$ is unsatisfiable then 
every model of $\T_0  \cup \Gamma$ is a model of  $\T_0 \cup \forall
y \Gamma_T(y)$. 

In \cite{Sofronie-lmcs2018} it is shown that if the extension 
${\mathcal T}_0 \subseteq {\mathcal T}_0 \cup
  {\mathcal K}$ satisfies condition $({\sf Comp}^{\Psi}_{f})$ then also the
  extension 
${\mathcal  T}_0 \cup \Gamma \subseteq {\mathcal T}_0 \cup \Gamma \cup   {\mathcal K}$
  satisfies condition $({\sf Comp}^{\Psi}_{f})$. If $\K$ is flat and linear then the
  extension is $\Psi$-local. 
Let $T = \Psi_{\K}(G) = \Psi({\sf est}(\K, G))$. 
By $\Psi$-locality, ${\mathcal T}_0 \cup \Gamma \cup {\mathcal K} \cup G$ is
unsatisfiable
if and only if ${\mathcal T}_0 \cup \Gamma \cup {\mathcal K}[T] \cup G$
is unsatisfiable, if and only if  (with the notations in Steps 1--5 of
Algorithm~1) 
${\mathcal T}_0 \cup \Gamma \cup {\mathcal K}_0 \cup G_0
\cup {\sf Con}_0 \cup {\sf Def}$ is unsatisfiable. 
Let ${\mathcal A}$ be a model of $\T_0  \cup \Gamma$. Then in ${\mathcal A}$ 
there are no possible values for the constants $G_1({\overline c_p},
{\overline c_f}, {\overline c}) = {\mathcal K}_0 \cup G_0
\cup {\sf Con}_0 \cup {\sf Def}$, for which $G_1({\overline c_p},
{\overline c_f}, {\overline c})$ is true in ${\mathcal A}$. 
Hence, ${\mathcal A} \not\models \exists {\overline x} G_1({\overline c_p},
{\overline c_f}, {\overline x})$, 
so (with the notation used when
describing Steps 1--5) 
${\mathcal A} \not\models \exists {\overline y}
\Gamma_2({\overline  y})$. 
It follows that ${\mathcal A} \models \forall  {\overline y} \Gamma_T({\overline  y})$. 
\QED

\medskip
\noindent This reduction method was implemented in {\sf sehpilot}
(for details cf.\  Section~\ref{tests}).

\section{Second-order quantifier elimination}
\label{second-order-qe}

\vspace{-1mm}
Let $\T$ be a theory with signature $\Pi = (S, \Sigma, {\sf Pred})$
and $P_1,\dots, P_n, Q_1,\dots, Q_{m}$ be predicate symbols which are
not in ${\sf Pred}$. 
Let $\Pi' =
(S, \Sigma, {\sf Pred} {\cup} \{ P_1,\dots, P_n \})$ and  
$\Pi'' = (S, \Sigma, {\sf Pred} {\cup} \{ Q_1,\dots, Q_m \})$; 
$F$ be a $\Pi'$-formula and $G$ a $\Pi''$-formula. 

\noindent 
A $\Pi$-structure $\A$ is a model of 
$\exists P_1 \dots P_n  ~ F$ (notation: $\A \models \exists P_1 \dots
P_n  ~ F$) if there exists a $\Pi'$-structure 
${\cal B}$ such that ${\cal B} \models F$ and ${\cal B}_{|_\Pi} = \A$.

\noindent We say that $\exists P_1 \dots P_{n} ~ F$ entails $\exists
Q_1 \dots Q_{m} ~ G$ w.r.t.\ $\T$ (and use the notation: 
$\exists P_1 \dots P_{n} ~ F \models_{\T} \exists
Q_1 \dots Q_{m} ~ G$) iff for every $\Pi$-structure ${\cal A}$ 
which is a model of $\T$,  
if $\A \models \exists P_1 \dots P_{n} ~ F$ then $\A \models \exists
Q_1 \dots Q_{m} ~ G$. 

\noindent If there exists a first-order formula $F_0$ 
over the signature $\Pi$ such that for every model ${\cal A}$ of
$\T$, ${\cal A} \models F_0$ iff 
${\cal A} \models \exists P_1 \dots P_n ~ F$, we say that $F_0$ and $\exists
P_1 \dots P_n ~ F$
are equivalent w.r.t.\ $\T$ (and write $F_0 \equiv_{\T} \exists P_1
\dots P_n ~ F$).

\smallskip
\noindent 
We consider here only the elimination of one predicate; for
formulae of the form $\exists
P_1 \dots P_n ~ F$ the process can be iterated.  
Let $\T$ be a theory with signature $\Pi = (S, \Sigma, {\sf Pred})$
and let $\Pi' = (S, \Sigma, {\sf Pred} \cup \{ P \})$, where
$P \not\in {\sf Pred}$. 

\smallskip
\noindent 
Let $F$ be a universal first-order $\Pi'$-formula. Our goal is to compute, if
possible, a first-order $\Pi$-formula $G$ such that $G \equiv_{\T} \exists
P~ F$. 
We adapt the hierarchical superposition calculus
proposed in \cite{BaumgartnerWaldmann13,BaumgartnerWaldmann2019}
to this case.

\medskip
\noindent We consider theories $\T$ over many-sorted signatures $\Pi = (S, \Sigma,
{\sf Pred})$, where the set of sorts $S = S_i \cup S_u$ consists of a 
set $S_i$ of interpreted sorts and a set $S_u$ of uninterpreted
sorts. The models of the theories are $\Pi$-structures $\A = (\{
A_{s}\}_{s \in S}, \{ f_{\A} \}_{f \in \Sigma}, \{ p_A \}_{p \in {\sf
    Pred}})$, where each support of interpreted sort is considered to be fixed. 
Following the terminology used in
\cite{BaumgartnerWaldmann13,BaumgartnerWaldmann2019}, 
we will refer to elements in the fixed domain of sort $s \in S_i$ as {\em
  domain elements of sort $s$}.

\noindent Let $F$ be a universal first-order formula over signature $\Pi' = (S,
\Sigma, {\sf Pred} \cup \{ P \})$. 
We can assume, without loss of generality, that $F$ is a set of
clauses of the form $\forall {\overline x} ~ D({\overline x}) \vee C({\overline
  x})$, where $D({\overline x})$ is a clause over the signature $\Pi$ 
and $C({\overline x})$ is a clause containing literals of the form 
$(\neg) P(x_1, \dots, x_n)$, where $x_1, \dots, x_n$ are variables% not necessarily distinct.
\footnote{We can bring the clauses to this
  form using variable abstraction.}.
Such clauses can also be represented as 
{\em constrained clauses} in the form 
$ \forall {\overline x} ~ \phi({\overline x})~ || ~C({\overline x}),
 \text{ where } \phi({\overline x}) := \neg D({\overline
  x}).$
We will refer to clauses of this form as constrained $P$-clauses. 

\smallskip
\noindent Let $\succ$ be a strict, well-founded ordering on terms that is
compatible with contexts and stable under substitutions. 
As in \cite{BaumgartnerWaldmann2019} we assume that
$\succ$ has the following properties:\footnote{These conditions are 
satisfied by an LPO with an operator precedence in which the predicate
symbol $P$ (which can be regarded as function symbol with output sort
${\sf bool}$) is larger than the other operators and domain elements are
minimal w.r.t.\ $\succ$ which is supposed to be well-founded on the domain elements.}
\begin{itemize}
\item[(i)] $\succ$ is total on ground terms, 
\item[(ii)] $t \succ d$ for every domain element $d$ of interpreted sort $s$
and every ground term $t$ 
that is not a domain element. 
\end{itemize}
Let $\mathit{HRes}^P_{\succ}$ be the calculus containing the following
ordered resolution and factorization rules for constrained
$P$-clauses: 

\

$\begin{array}{ll}
\displaystyle{{\phi_1~ ||~ P({\overline x}) \vee C \quad \quad \phi_2 ~ || ~
\neg P({\overline y}) \vee D} 
\over 
{(\phi_1 \wedge \phi_2)\sigma~||~(C \vee D)\sigma}}  \quad \quad \quad \quad \quad \quad \displaystyle{{\phi~ ||~
P({\overline x}) \vee  P({\overline y}) \vee C} 
\over 
{\phi\sigma ~ || ~ (P({\overline x}) \vee C )\sigma}}
\end{array}$

\smallskip
\noindent \begin{tabular}{@{}rlll}
\text{ where} (i)&  $\sigma = {\sf mgu}(P({\overline x}), P({\overline
  y}))$  & (i) & $\sigma = {\sf mgu}(P({\overline x}), P({\overline
    y}))$\\
 (ii) & $P({\overline x}) \sigma$  is strictly maximal in
$(P({\overline x}) \vee C)\sigma$ & (ii) & $P({\overline x}) \sigma$  is
maximal in \\
(iii) & $\neg P({\overline y})\sigma$ is maximal in  $(\neg
P({\overline y}) \vee D) \sigma.$ & & $(P({\overline x}) \vee C)\sigma$ 
\end{tabular}

\

\

\noindent {\bf Redundancy.} The inference rules are supplemented by a
redundancy criterion ${\cal R} = ({\cal R}_c, {\cal R}_i)$ meant to
specify:
\begin{itemize}
\item a set ${\cal R}_c$ of redundant clauses (which can be removed),
  and 
\item a set ${\cal R}_i$ of redundant inferences (which do not need to be computed). 
\end{itemize}
We say that a set of clauses $N^*$ is saturated up to
${\cal R}$-redundancy w.r.t.\ $\mathit{HRes}^P_{\succ}$ if every 
$\mathit{HRes}^P_{\succ}$ inference with premises in $N^*$ is
redundant (i.e.\ in ${\cal R}_i$).

\medskip
\noindent The following notion of redundancy ${\cal R}^0_c$ for clauses is often used: A (constrained) clause is
redundant w.r.t.\ a set $N$ of clauses if all its ground instances are entailed w.r.t.\ $\T$
by ground instances of clauses in $N$ which are strictly smaller w.r.t.\
$\succ$. We will use the following notion of redundancy for
inferences: If ${\cal R}_c$ is a redundancy criterion for clauses, we say that an 
inference $\iota$ on ground clauses is redundant w.r.t.\ $N$ if either one of
its premises is redundant w.r.t.\ $N$ and ${\cal R}_c$ or, if $C_0$ is the conclusion
of $\iota$ then there exist clauses $C_1, \dots ,C_n \in N$ that are
smaller w.r.t.\ $\succ$ than the maximal premise of $\iota$ and
$C_1,\dots,C_n \models C_0$.

\medskip
\noindent 
A non-ground inference is redundant if all its ground instances are redundant.  

\begin{ex}[Semantic $\T$-entailment; redundancy criterion ${\cal R}_{\T}$]
{\em 
We say that a constrained $P$-clause $\forall {\overline x} ~ (\phi({\overline x}) ~||~
C({\overline x}))$ is {\em $\T$-semantically entailed} by 
$\forall {\overline x} ~ (\psi({\overline y}) ~||~ D({\overline
  y}))$ if the following conditions hold: 
\begin{itemize}
\item[(i)] $C = D$, 
\item[(ii)] $\T \models \forall {\overline x} ~ \phi({\overline x}) \rightarrow
\psi({\overline x})$ and 
\item[(iii)] 
$\neg \phi({\overline x})\sigma \succ \neg \psi({\overline y})\sigma$ 
for every ground substitution $\sigma$. 
\end{itemize}
We say that a clause $C$ is {\em ${\cal R}_{\T}$-redundant w.r.t.\ a set $N$ of
clauses} if it is $\T$-semantically entailed by a clause in
$N$. 

\noindent 
Note that if $C_1 {=} (\phi({\overline x}) || C({\overline x}))$ is
$\T$-semantically entailed by $C_2 {=} (\psi({\overline y}) || D({\overline y}))$
then $C_1 \sigma \succ C_2 \sigma$ and $C_2 \sigma \models_{\T} C_1\sigma$ 
for every ground substitution $\sigma$, so ${\cal R}_{\T}$-redundant clauses are ${\cal
  R}^0_c$-redundant.

\noindent We call the notion of redundancy induced on
inferences also ${\cal R}_{\T}$-redundancy. 
} \QEX
\label{redundancy-def}
\end{ex}
Let ${\cal R} = ({\cal R}_c, {\cal R}_i)$ be a redundancy
criterion with ${\cal R}_c \subseteq {\cal R}^0_c$. 
We want to prove that if $N$ is a set of constrained $P$-clauses over background theory
$\T$, $N^*$ its saturation (up to ${\cal R}$-redundancy) under
$\mathit{HRes}^P_{\succ}$, and $N^*_0$ the
set of clauses in $N^*$
not containing $P$, then for every model $\A$ of $\T$,
$\A$ is a model of $N^*_0$ if and only if there exists a 
$\Pi'$-structure ${\cal B}$ with $\B \models
  N$ and ${\cal B}_{|_\Pi} = {\cal A}$.
The proof of this fact is very similar to the
proof of the completeness of hierarchical superposition. Since our
goal is different, we present here all the details just for the sake
of completeness. (The results are probably known, already in
\cite{BGW94} it was mentioned that hierarchical superposition can be 
used for second-order quantifier elimination.)
We start with a lemma. 

\begin{lem}
Let $\Pi = (S, \Sigma, {\sf Pred})$, $\T$ be a theory with signature
$\Pi$ and let 
$\A = (\{ A_s \}_{s \in S}, \{ f_{\A} \}_{f \in \Sigma}, \{ p_{\A}
\}_{p \in {\sf Pred}})$ be a $\Pi$-structure which is a model of $\T$. 
For every element $a \in A_s$ we add a new constant  of sort $s$
(which we denote $a$). 
Let $C_A {=} \bigcup_{s \in S} A_s$ be the set of all constants
introduced this way,  and ${\cal A}^A$ be the extension of ${\cal
  A}$ with constants from $C_A$ which are interpreted in the usual
way. 

\noindent Let $N$ be a set of clauses over signature $\Pi$. 
Then the following are equivalent: 
\begin{itemize}
\vspace{-1mm}
\item[(1)] ${\cal A}$ is a model of $N$. 
\item[(2)] ${\cal A}^A$ is a model of the set $N_A$ of all ground instances
of $N$ in which the variables are replaced with constants in $C_A$. 
\vspace{-1mm}
\end{itemize}
\label{lemma:instances}
\end{lem}
{\em Proof:} 
(1) $\Rightarrow$ (2):  Assume that ${\cal A}$ is a model of $N$. 
Let $C$ be a clause in $N_A$. Then $C$ is obtained from a clause $C'
\in N$ by replacing every variable $x$ with a constant $a_x \in C_A$
(such that if the variable $x$ has sort $s$ then $a_x\in A_s$). 
Let $\beta : X \rightarrow {\cal A}$ be defined by $\beta(x) = a_x$
for every $x \in X$ occurring in $C'$ and defined arbitrarily for all
the other variables. Since ${\cal A}$ is a model of $N$, 
the clause $C'$ is true in ${\cal A}$ in the valuation $\beta$. 
But ${\cal A}(\beta(C'))$ is obtained by evaluating the function and
predicate symbols as in ${\cal A}$ and every variable $x$ occurring in
$C'$ as $a_x$. This is exactly the value of $C'$ in ${\cal A}^A$, thus
${\cal A}^A$ is a model of $C$. 

\smallskip
\noindent (2) $\Rightarrow$ (1): Assume that ${\cal A}^A$ is a model of $N_A$.
Let $C \in N$ and let $\beta : X \rightarrow {\cal A}$ be
a valuation. For every $x \in X$, let $a_x := \beta(x)$. 
As discussed before, the value of $C$ in ${\cal A}$ w.r.t.\ $\beta$ 
is the same as the value of $C' = C \sigma$ in ${\cal A}^A$, 
where $\sigma$ is the substitution that associates with every variable
$x$ the constant $a_x \in C_A$. Since ${\cal A}^A$ is a model of the
set $N_A$ and $C' \in N_A$, ${\cal A}^A$ is also a model of $C'$. 
It follows that $C$ is true in ${\cal A}$ w.r.t.\ $\beta$ for every
valuation $\beta$, i.e.\ that  ${\cal A}$ is a model of $N$. \QED

\begin{thm}
Let $N$ be a set of constrained $P$-clauses over background theory
$\T$, $N^*$ its saturation (up to ${\cal R}$-redundancy) under
$\mathit{HRes}^P_{\succ}$, and $N^*_0$ the
set of clauses in $N^*$
not containing $P$. For every model $\A$ of $\T$ the
following are equivalent: 
\begin{itemize}
\item[(1)] $\A$ is a model of $N^*_0$. 
\item[(2)] There exists a $\Pi'$-structure ${\cal B}$ with $\B \models
  N$ and ${\cal B}_{|_\Pi} = {\cal A}$.
\end{itemize}
\label{thm:model-constr}
\label{thm:redundancy}
\end{thm}
{\em Proof:} 
First, note that for constrained $P$-clauses the hierarchical
superposition calculus specializes to $\mathit{HRes}^P_{\succ}$: 
With the terminology used in \cite{BGW94,BaumgartnerWaldmann13,BaumgartnerWaldmann2019}, 
the background signature is $\Pi$; the only foreground symbol 
is $P$. Since there are no ``background''-sorted  
terms starting with a ``foreground'' function symbol, sets of $P$-clauses are sufficiently complete. 
(Even if we regard predicates as functions with values in the domain
$\{0, 1\}$, since predicates can only take values 0 or 1 sufficient
completeness is guaranteed.) 
Since in the special case we consider there are no foreground terms,
in this case all substitutions are simple as well.
 
\medskip
\noindent (2) $\Rightarrow$ (1) follows from the soundness of the
hierarchical superposition calculus. 

\medskip
\noindent (1) $\Rightarrow$ (2) is proved with a model construction similar
to the one used for proving completeness of hierarchical
superposition. Let $\A$ be a  model of $N^*_0$ and $\T$.
Consider the extension $\Pi^A$ of the signature $\Pi$ obtained by adding to
$\Sigma$ a set $C_{\A}$ containing a constant $a$ of sort $s$ for every 
element $a \in A_s$.

Since $\A$ is a model of $N^*_0$, by Lemma~\ref{lemma:instances}, 
$\A^A$ is a model of the set ${N^*_0}_A$ of instances of $N^*_0$ 
in which the variables are replaced with constants in $C_A$.

By Zermelo's theorem, there exists a total well-founded strict order $\succ$
on the set of all constants in $\Pi$, and starting with this ordering
we can obtain a total well-founded strict ordering (which we denote
again with $\succ$) on the set of all ground terms over $\Pi^A$, 
which can be extended in the usual way 
to the set of all ground clauses over the signature ${\Pi'}^A =
(S, \Sigma \cup C_A, {\sf Pred} \cup \{ P \})$, such that the literals 
containing the predicate symbol $P$ are larger than the literals not
containing $P$.\footnote{This is compatible with regarding $P$ as a
  function symbol with output sort ${\sf bool}$ and using an ordering in
  which  $P$ is the largest function symbol.}

Consider the clauses in the set $N^*_A$ ordered increasingly according
to the clause ordering induced by the atom ordering. Since $N^*$ is
saturated w.r.t.\ $\mathit{HRes}^P_{\succ}$, $N^*_A$ is also saturated 
w.r.t.\ $\mathit{HRes}^P_{\succ}$. 
We construct a model for $N^*_A$ using a canonical model
construction, similar to the one usually used for proving completeness
of ordered resolution. We sketch the construction here: 

We start with an interpretation in which all atoms in the set 
$\{ P(a_1, \dots, a_k) \mid i \in \{ 1, \dots n \}, a_j
\in C_A \text{ constant of suitable sort, for all } j \in \{ 1, \dots, k \} \}$
are false. 
The clauses in $(N^*_0)_A$ are smaller than the clauses containing the
predicate symbol $P$ and by assumption are all true in ${\cal A}$,
hence also in the interpretation that we start with (and will remain
true in the process of constructing $\B$). 
We therefore only need to consider the set $N^*_P$ of 
all constrained $P$-clauses containing the predicate symbol $P$. 
When considering a clause $C$, we assume that we already constructed a
partial interpretation ${\cal B}_{\prec C}$ that makes true all
clauses strictly smaller than $C$. 
\begin{itemize}
\item If $C$ is true in the partial interpretation ${\cal B}_{\prec C}$
nothing needs to be done. 
\item If $C$ is false  in the partial interpretation ${\cal B}_{\prec C}$ we
need to change  ${\cal B}_{\prec C}$ such that $C$ becomes true
(such that the clauses smaller than $C$ remain true). 
\end{itemize}
We proceed as follows: 
If $C$ is false in ${\cal B}_{\prec C}$ and contains exactly one
maximal literal which is positive (which needs to start with $P$ and
is for instance of the form $P({\overline a})$), 
we change the interpretation of $P$ such that it contains 
the tuple ${\overline a}$, i.e. such that $P({\overline a})$ becomes
true. We denote this by setting $\Delta_C := \{ P({\overline a})
\}$. Otherwise we do not change the interpretation, i.e.\  $\Delta_C
:= \emptyset$. 

\noindent The candidate model is the limit of all these changes, 
${\cal B} := {\cal A} \cup \bigcup_{C \in N_1} \Delta_C$. 

We can show that the expansion ${\cal B}^{A}$ of ${\cal B}$ with the constants in
$C_A$ (with the usual interpretation) is a model of $(N^*_P)_A$ 
in the usual way: Assume that there exists a clause in $(N^*_P)_A$
which is not true in ${\cal B}^{A}$. Since the ordering on $(N^*_P)_A$ is 
well-founded, we consider without loss of generality the smallest 
clause in $(N^*_P)_A$ which is false in $\B^A$. 
We can show in the usual way that using a resolution or a
factorization step we can produce a smaller clause false in $\B^A$, 
which is either in  $(N^*_P)_A$ or in $(N^*_0)_A$, so in both cases we
obtain a contradiction.  

Since $\B^A$ is a model of $N^*_A$, and $\B_s = \A_s$ for every sort
$s$ (hence $\B^A = \B^B$), it follows that $\B$ is a model of
$N^* = N^*_0 \cup N^*_P$. 
\QED

\

\begin{cor}
Let $\T$ be a theory, $N$ be a set of constrained $P$-clauses and
$N^*$ be a set of constrained $P$-clauses obtained by saturating $N$ in 
$\mathit{HRes}^P_{\succ}$ up to redundancy.  Let $\A$ be a
$\Pi$-structure which is a model of $\T$. Then  $\A$ is a model of 
$N^*_0$ if and only if there exists a $\Pi'$-structure ${\cal B}$ with 
$\B \models N$ and ${\cal B}_{|_\Pi} = {\cal A}$.
\end{cor}

\noindent {\em Proof:} Follows from the fact that in the proof of 
Theorem~\ref{thm:model-constr} the clauses which are redundant
are entailed (w.r.t. $\T$) by clauses that are smaller hence cannot 
be minimal counterexamples and cannot influence the way 
model $\B$ is built because every redundant clause $C$ is true 
in $\B_{\prec C}$. \QED

\subsection{Case 1: Saturation is finite} 
If the saturation $N^*$ of $N$ under
$\mathit{HRes}^P_{\succ}$ (up to ${\cal R}$-redundancy) is finite and 
$N^*_0$ is the set of clauses in $N^*$ not containing $P$ then, 
by Theorem \ref{thm:model-constr}, the universal closure of the conjunction of
the clauses in $N^*_0$ is  equivalent to $\exists P~ N$.

\begin{ex}
{\em 
Consider a class of graphs described by the following set $N$ of constrained
$E$-clauses: 

\smallskip
\noindent $\{ (1) \pi^i(u,v)  || E(u,v), ~~ (2) \pi^t(u,w,v)  || E(u,v)
\rightarrow E(u,w),~~ 
(3) \pi^e(u,v)  || \lnot E(u,v) \}$ 

\smallskip
\noindent 
For arbitrary predicates $\pi^i, \pi^e$ and $\pi^t$ we can generate
with $\mathit{HRes}^E_{\succ}$ an infinite set of clauses including,
e.g.,  
all clauses of the form: 

\smallskip
\noindent $
(4_n)~ \pi^i(u,v) \land \pi^t(u,w_1,v) \land
\pi^t(u,w_2,w_1) \land \dots \land \pi^t(u, w_n, w_{n-1})  ~ || ~ E(u,w_n) 
$ 

\smallskip
\noindent 
If we assume that $\pi^i, \pi^e, \pi^t$ satisfy the additional
 axioms defining a theory $\T_{\pi}$: 

\smallskip
$\begin{array}{lrrll}
(c1) & \forall u, v, w ~ ~~ & \pi^i(u, v) \land \pi^t(u, w, v) &
\rightarrow & \pi^i(u,w) \\
(c2) & \forall u, v, w ~ ~~ & \pi^e(u, w) \land \pi^t(u, w, v) &
\rightarrow & \pi^e(u,v) \\
(c3) & \forall u, v, w, x ~~ & \pi^t(u,w,v) \land \pi^t(u,x,w) & \rightarrow & \pi^t(u,x,v) \\
\end{array}$

\smallskip
\noindent then all inferences by resolution between clauses (1) and
(2), (2) and (3) are $\T_{\pi}$-redundant. 
The inferences between (2) and (2) are also $\T_{\pi}$-redundant: 
Consider a ground instance of such an inference (the maximal literals
are underlined): 
\[ \pi^t(c_1,c_2,c_3) || \underline{E(c_1,c_3)} \rightarrow E(c_1,c_2)
\quad \pi^t(c_1,c_3,c_4) || E(c_1,c_4) \rightarrow
\underline{E(c_1,c_3)}
\over 
\pi^t(c_1,c_2,c_3),  \pi^t(c_1,c_3,c_4) || E(c_1,c_4) \rightarrow
E(c_1,c_2) \]
The ground instance $C$ of (2) $\pi^t(c_1,c_2,c_4) || E(c_1,c_4) \rightarrow E(c_1,c_2)$
differs from the conclusion of the inference above only in the
background part and is smaller
than the first premise of the inference.
If $(c3)$ holds then $C$ entails
the conclusion of the inference above, which makes the inference
redundant w.r.t.\ $\T_{\pi}$.

\noindent Thus, only the inference of clauses (1) and (3) 
yields a non-redundant resolvent:  

\smallskip
$(4) \quad  \pi^i(u,v, r_1) \land \pi^e(u,v, r_2) || \bot$ 

\smallskip
\noindent so $N^* = N \cup \{ (4) \}$ is saturated up to $\T_{\pi}$-redundancy. By
Theorem~\ref{thm:model-constr}, $N$ is
satisfiable iff  $\forall u, v (\pi^i(u,v) \land \pi^e(u,v)
\rightarrow \bot)$ is satisfiable w.r.t.\ $\T_{\pi}$. 
} \QEX
\label{ex-sat1}
\end{ex}
When modelling concrete situations, the
predicates $\pi^i, \pi^e$ and $\pi^t$ might not be arbitrary, but
might have definitions using other symbols with given properties.

\begin{ex}
{\em 
The theory
$\T_{\pi}$ might be actually described in a detailed way. 
Let $\class{C}(r_1, r_2)$ be a graph class described by the set $N$ of
axioms in Example~\ref{ex-sat1}, where 
$\pi^i, \pi^e, \pi^t$ are defined by the axioms: 

\medskip
$\begin{array}{ll} 
{\sf Def}_{\pi}(r_1, r_2)  = \{ & \forall u, v~  
\pi^i(u,v) \leftrightarrow u {\neq} v \land d(u,v) {\leq} r_1(u),\\
& \forall u, v ~ \pi^e(u, v) \leftrightarrow  d(u,v) {>} r_2(u), \\
& \forall u, v, w ~ \pi^t(u,w,v) \leftrightarrow u {\neq} w \land d(u,w)
{\leq} d(u,v) \},
\end{array}$ 

\medskip
\noindent where $d$ is a distance or cost function.  
We can regard the theory extension $\T \subseteq \T_{\pi} = \T \cup
{\sf Free}(r_1, r_2) \cup {\sf Def}_{\pi}$, where $\T$ 
is one of the theories  $\T^u_d$, $T^p_d$, $\T^n_d$ or $\T^m_d$ introduced in
Theorem \ref{thm:distance}, and ${\sf Free}(r_1, r_2)$ the theory in
which $r_1, r_2$ are regarded as uninterpreted unary function symbols.
Therefore, $\T$ can be represented as a $\Psi$-local extension of the disjoint
combination  of a theory of real numbers and of pure equality, for a
suitable closure operator. 
We can use the hierarchical reduction in Theorem \ref{lemma-rel-transl} to check that 
$(c1), (c2)$ and $(c3)$ are valid w.r.t.\ $\T$. 
}\QEX
\label{ex-sat2}
\end{ex}
In applications we might not be interested in 
checking the satisfiability of $N$ or the satisfiability of $\forall u, v (\pi^i(u,v) \land \pi^e(u,v)
\rightarrow \bot)$ w.r.t.\ $\T_{\pi}$, but in a 
{\em specific model} $\A$ satisfying $\T_{\pi}$ (we refer to it as
``canonical model''). 

\smallskip
\noindent This is the case, for instance, in the applications in wireless
network theory analyzed in Section~\ref{networks}: The vertices of the
graphs considered in this context are very often points in the
Euclidian space, 
and the distance is a concrete function which can be, for instance, 
the Euclidean metric, or a concrete cost function -- which might 
satisfy additional properties  (for instance positivity or symmetry).
If we want to analyze such graph classes in full generality, we might 
assume that some of the properties of some of the parameters are 
not fully specified.

\medskip
\noindent 
Let $\A$ be a  model of a theory $\T$ describing properties of
function symbols in a set $\Sigma$ we want to model.
We assume that $\Sigma$ contains a set of 
``parameters'' $\Sigma_{\sf par}$ (function symbols whose properties are
``underspecified'' in $\T$). 
In some situations, if we are given a set $N$ of constrained clauses, 
we might be interested in obtaining (weakest) universal conditions $\Gamma$ on 
$\Sigma_{\sf par}$ such that 
for {\em every} fixed model $\A$ of $\T$ which also satisfies $\Gamma$,  
there exists an interpretation for $P$ in $\A$ for which $N$ is
satisfied, i.e.\ $\A \models \exists P ~ N$. We present a situation in
which this is possible.

\begin{thm}
Let $\T$ be a theory with signature $\Pi = (S, \Sigma, {\sf Pred})$,  
$N$ a set of constrained $P$-clauses. Assume that the saturation $N^*$ 
of $N$ up to $\T$-redundancy w.r.t.\
$\mathit{HRes}^P_{\succ}$ is finite; let $N^*_0$
be the set of clauses in $N^*$ not containing $P$. 

\noindent Let $\Sigma_{\sf par} \subseteq \Sigma$ be a set of parameters.
Assume that one of the following conditions holds: 
\begin{itemize}
\item[(i)] $\T$ allows quantifier elimination or 
\item[(ii)] $\T_0 \subseteq \T =
\T_0 \cup \K$ is a local theory
extension satisfying condition $({\sf Comp}^\Psi_f)$ for a suitable
term closure operator $\Psi$ and $\T_0$ allows
quantifier elimination.
\end{itemize}
If (i) holds, we can use quantifier elimination and if (ii) holds 
then we can use Algorithm~1 
to obtain a (weakest) universal constraint $\Gamma$ on the parameters such that
{\em every}
model $\A$ of $\T \cup \Gamma$ is a model of (the universal closure
of) $N^*_0$, hence $\A \models \exists P ~ N$.
\label{thm:symbol-elim}
\end{thm}
{\em Proof:}  
By the completeness of 
the hierarchical superposition calculus, $N$ is satisfiable 
iff the set $N^*_0$ of clauses in $N^*$ which do not contain $P$ is
satisfiable.  We denote by $\forall  {\overline x} N^*_0({\overline
  x})$
the formula represented by the set of clauses $N^*_0$.

\noindent Let $\A$ be a model of $\T$. Assume that $\A$ is not a model for
$\forall {\overline x} N^*_0({\overline x})$. 
Then $\exists {\overline x} \neg N^*_0({\overline x})$ is true in $\A$. 
In particular, it follows that $\T \wedge \exists {\overline x}
\neg N^*_0({\overline x})$ is satisfiable. 
We can apply Algorithm~1 to construct a weakest universal formula
$\Gamma$ over the signature $\Sigma_{\sf par}$ 
with the property that $\T \cup \Gamma \cup \exists {\overline x}
\neg N^*_0({\overline x})$ 
is unsatisfiable, i.e.\ with the property that 
$\T \cup \Gamma \models \forall {\overline x} N^*_0({\overline x})$.
\noindent Then every model $\B$ of $\T$ which also satisfies the constraints in
$\Gamma$ is a model of $\forall {\overline x} N^*_0({\overline x})$. \QED

\begin{ex}
{\em 
Consider again the situation described in Example~\ref{ex-sat2}.
We show how one can use Theorem~\ref{thm-alg-1}
and Algorithm 1
to derive  constraints $\Gamma$ on the parameters
$r_1, r_2$ under which for every model $\A$ of $\T^u_d \cup \Gamma$ 

\medskip
$\pi^i(u,v) \wedge \pi^e(u,v) = (u \neq v \land d(u,v) \leq r_1(u))
\wedge (d(u, v) > r_2(u))$

\medskip
\noindent 
is unsatisfiable in $\A$
(we consider the case in which $d$ is an uninterpreted function; 
other axioms for $d$ can be analyzed as well).

\noindent Note that the formula above is unsatisfiable in any model $\A$
of $\T^u_d$ whose support $A_p$ of sort ${\sf p}$ has cardinality 1.
If we only consider models $\A$ with $|A_p| \geq 2$ then we can proceed as follows:

\smallskip
\noindent 
{\bf Step 1:} We purify the formula by introducing new constants: $c_d := d(u,
v), c_1 = r_1(u), c_2 = r_2(u)$ and obtain:  $(u \neq v \land c_d \leq c_1 \land c_d > c_2)$.

\smallskip
\noindent {\bf Step 2:} We quantify existentially all constants not
denoting terms starting with -- or used as arguments of -- $r_1, r_2$
and obtain: $\exists v, \exists c_d (u \neq v \land c_d \leq c_1 \land c_d > c_2)$.

\smallskip
\noindent 
{\bf Step 3:} 
After 
quantifier elimination in a combination of ${\sf LI}({\mathbb R})$ and
the theory of sets with cardinality $\geq 2$ with equality
\cite{PeuterSofronie2019} we obtain $\Gamma_1(c_1, c_2, u): c_2 <
c_1$.\footnote{We can consider only models $\A$ whose support of sort
  ${\sf p}$ is infinite. The theory that formalizes this is the model
  completion of the theory  ${\cal E}$ of pure equality which allows
  quantifier elimination. We can then use the method for quantifier
  elimination in combinations of theories with QE described in \cite{PeuterSofronie2019}.}

\smallskip
\noindent 
{\bf Step 4:} We replace the constants $c_1, c_2$ with the terms they
denote and quantify the arguments existentially and obtain: 
$\exists u (r_2(u) < r_1(u))$.

 \smallskip
\noindent 
{\bf Step 5:} We negate this condition and obtain: $\forall u (r_1(u)
\leq r_2(u))$. 
\label{ex:sat}
}\QEX \end{ex}

\begin{ex}
{\em We find an axiomatization for the graph class
$\class{C}^- = \{ G^- \mid G \in \class{C} \}$, when class $\class{C}$
is described by the set $N$ of constrained
clauses in Example~\ref{ex-sat1} and $\pi^i, \pi^e$ and $\pi^t$ satisfy conditions $(c1), (c2),
(c3)$. Let $N^* = N \cup \{ (4) \}$ be obtained by saturating 
$N$ under $\mathit{HRes}^E_{\succ}$ up to redundancy. 
A graph $H = (V, F) \in \class{C}^-$ iff there exists a graph $G =
(V,E) \in \class{C}$ such that $H = G^-$. 
This condition can be described
by $M = N^* \cup {\sf Tr}(E, F)$, where 
${\sf Tr}(E,F) = \{ \forall x, y ~ (F(x, y) \leftrightarrow E(x, y) \wedge E(y,x)) \}$, 
which can be written in the form of constrained clauses as: 

\medskip
\noindent ${\sf Tr}(E, F) ~ = ~ \{  F(x, y) || E(x, y), ~  F(x, y) || E(y,
x),  ~ \neg F(x, y) || \neg E(x, y) \lor \neg E(y, x) \}
$ 

\smallskip
\noindent To find an axiomatization for the class $\{ (V, F)
\mid \exists (V, E) \in \class{C} \text{ with } (V, F) = (V, E)^- \}$
we need to eliminate the second-order quantifier from 
the formula $\exists E (N^* \cup {\sf Tr}(E, F))$. 

\medskip
\noindent The base theory is  $\T \cup \mathit{UIF}_F$, the extension
of $\T$ with the uninterpreted function symbol $F$, with signature $\Pi_F = (\Sigma,
{\sf Pred} \cup \{ F \})$.

\noindent Since the background theory in this case is not arithmetic, 
and since the method for second-order quantifier elimination
implemented in SCAN \cite{Gabbay-Ohlbach} is very similar to
$\mathit{HRes}^E_{\succ}$, 
we used SCAN on the clause set $N^-$:

\smallskip
$\begin{array}{llrl}
(c1)~~~~ & \forall u, v, w, x & \pi^t(u,w,v) \wedge \pi^t(u,x,w) & \rightarrow \pi^t(u,x,v)\\
(c2) & \forall u, v, w  & \pi^i(u,v) \land \pi^t(u,w,v) & \rightarrow \pi^i(u,w)\\
(c3) & \forall u, v, w  & \pi^e(u,w) \land \pi^t(u,w,v) & \rightarrow \pi^e(u,v)\\[2ex]
(1) & \forall u, v    & \pi^i(u,v) & \rightarrow E(u,v) \\
(2) & \forall u, v    & \pi^e(u,v) & \rightarrow  \neg E(u,v) \\
(3) & \forall u, v, w &  \pi^t(u,w,v) \land E(u,v) & \rightarrow
E(u,w)\\
(T1) & \forall u, v   &  F(u,v) & \rightarrow E(u,v) \\
(T2) & \forall u, v   &  F(u,v) & \rightarrow E(v,u) \\
(T3) & \forall u, v   &  F(u,v) \wedge E(v,u) & \rightarrow F(u,v)
\end{array}$

\smallskip
\noindent and obtained a set of clauses representing the 
formula containing axioms $N_{\T} = \{ (c1), (c2), (c3), (4) \}$,
where 

\smallskip
$(4)~~~~ \forall u, v ~ \pi^i(u, v) \wedge \pi^e(u, v) \rightarrow
\perp$   

and axioms $N^-_F = \{ (F_1), \dots, (F_6) \}$: 

 \smallskip
$\begin{array}{llrl}
(F_1) & \forall x, y &  F(x,y) & \rightarrow F(y,x)\\
(F_2) & \forall x, y & F(y,x) & \rightarrow \neg \pi^e(x,y) \\  
(F_3) & \forall x, y & F(x,y) & \rightarrow \neg \pi^e(x,y) \\ 
(F_4) & \forall x, y &  \pi^i(x,y) \land \pi^i(y,x) & \rightarrow
F(y,x) \\
(F_5) & \forall x, y, z & \pi^i(x,y) \land \pi^t(y,x,z) \land F(y,z)
& \rightarrow F(x,y) \\ 
(F_6) & \forall x, y, z, u ~~&  \pi^t(x,y,z) \land F(x,z) \land \pi^t(y,x,u)
\land F(y,u) & \rightarrow F(y,x) 
\end{array}$

\medskip
\noindent The universal closure $G$ of the conjunction of these clauses 
is equivalent w.r.t.\ $\T \cup \mathit{UIF}_{F}$ to
the formula $\exists E (N^* \cup {\sf Tr}(E, F))$, and thus
axiomatizes $\class{C}^-$. 
\label{ex:trans-minus}

} \QEX

\end{ex}

\subsection{Case 2: Finite representation of possibly infinite
  saturated sets}

The saturation of a set $N$ of constrained $P$-clauses up to redundancy under $\mathit{HRes}^P_{\succ}$ 
might be infinite. 
We here consider a very special case 
under which a finite set of constrained $P$-clauses 
$N = \{ \phi_i({\overline x}) ~ || ~
C_i({\overline x}) \mid i  = 1, \dots, n \}$
can have a saturation that can be finitely described: 
The situation in which the set of
clauses $\{ C_1, \dots, C_n \}$ can be finitely saturated under
ordered resolution.   
\begin{thm}
Let $N = \{ \phi_i({\overline x})  || 
C_i({\overline x}) \mid i  = 1, \dots, n \}$ be a finite set of
constrained $P$-clauses and $N_P = \{ C_1, \dots, C_n \}$.

Assume that the saturation of $N_P$ under ordered resolution
is finite, $N_P^* = \{ C_1, \dots, C_n, C_{n+1}, \dots, C_{n+k}
\}$, and the set ${\cal I}_P$ of all possible inferences used for deriving these
clauses is finite and can be effectively described. 
If $\perp \not\in N_P^*$, then $\exists P~ N \equiv_{\T}
  \top$.
Assume now that $\perp \in N_P^*$. 
Let $\A$ be a model of $\T$, $\T_{\A}$ the theory with $\A$ as
canonical model (i.e.\ $\T_{\A} = Th(\A)$). 
Let $N_A$ be
  the set of all instances of $N$ in which the variables are
  replaced with elements in $\A$ (seen as constants). Then:

\begin{itemize}
\item[(1)]  The saturation $N^*_A$ of  $N_A$ up to ${\cal R}_{\T}$-redundancy 
can be described as 
$ N^*_A = \{ \mu^{\A}_i({\overline a}) ~ || ~
C_i({\overline a}) \mid i  = 1, \dots, n + k, {\overline a} \text{
  elements of } \A\}, $
where $(\mu^{\A}_i)_{i = 1, n+k}$ are given by the minimal model of
the constrained Horn clauses\footnote{The definitions are presented in
  Appendix~\ref{app:constrained-Horn-Clauses}.}
 $CH_N$ w.r.t.\ $\T_{\A}$:

\smallskip
\noindent 
{$\begin{array}{rl} 
CH_N = \{ & \phi_i({\overline x}) \rightarrow  \mu_i({\overline x})
 \mid  i \in \{
1, \dots, n \}\} \\
\cup \{& (\mu_i({\overline x}) \wedge \mu_j({\overline y}))\sigma \rightarrow
\mu_k({\overline z})  \mid C_k({\overline z}) \text{ is obtained by a
    resolution }  \\
& \hspace{1.5cm} \text{ inference in } {\cal I}_P\text{ from } C_i({\overline x}) \text{ and } C_j({\overline y}) \text{ with m.g.u.\ } \sigma \}\\
\cup \{& \mu_i({\overline x})\sigma  \rightarrow \mu_k({\overline z})
\mid  C_k({\overline z}) \text{ is obtained by a 
    factorization inference  }  \\
&  \hspace{1.5cm} \text{ in } {\cal I}_P \text{ from } C_i({\overline x}) \text{ with m.g.u.\ } \sigma \}.
\end{array}$
}

\item[(2)] Let $A^{\mu}$ be the extension of $\A$ with
  predicates $(\mu_i)_i$ whose interpretation is given by 
  $(\mu^{\A}_i)_i$. 
Let $j$ be such that $\perp = C_j$. Then 
$\A \models \exists
  P \, N$ 
iff 
$CH_N {\cup} \{ {\neg} \mu_j({\overline  x}) \}$ is satisfiable
w.r.t.\ $\T_{\A}$.
\end{itemize}
\label{thm:min-model}
\end{thm}
 {\em Proof:} Obviously, $CH_N$ is satisfiable (it has one
trivial model, in which all predicate symbols $\mu_i$ are true). 
If $\A$ is a model of $\T$, in this theorem we consider constrained
Horn clauses over an assertion language that has one canonical model,
namely $\A$, i.e.\ w.r.t.\ the corresponding theory $\T_{\A}$.  
In \cite{McMillanRybalchenkoBjorner} it is shown --
using a canonical model construction -- that every set $H$ of 
constrained Horn clauses 
over an assertion language that has canonical models has a  unique least
model. \\
This model is 
defined inductively by taking 

\smallskip
\noindent $I_0 {:=} \emptyset$ and 

\noindent  $I_{i+1} {:=} \{ r({\overline a}) {\mid}  {\sf body}(x) {\rightarrow} r(x) \in H,  I_i
  \models_{\A} {\sf body}({\overline a}), {\overline a} \text{ is a
    tuple of constants in } \A \}$. 

\smallskip
\noindent The construction stabilizes
at the first limit ordinal $\omega$ with an interpretation
$I_{\omega}$; so the set of constrained Horn clauses $CH_N$ has a 
unique least model w.r.t.\ $\T_{\A}$. 

\medskip
\noindent This construction of this unique least model parallels the 
saturation process for the set of 
ground instances of the clauses $N^*_A$; the saturated set is: 

\smallskip
$N^*_A = \bigcup_{i = 1}^n \{ \xi^{k}_i({\overline
  a})||C_i({\overline a}) \mid k {\in} {\mathbb
  N}, {\overline a} \text{ is a sequence of elements in } \A \}$

\smallskip
\noindent 
where $\xi^{k}_i({\overline
  a})||C_i({\overline a})$ are clauses in $N_A$ or are obtained (in a
finite number of steps) from clauses in $N_A$ using resolution and/or
factorization. 
If we allow for potentially infinite disjunctions it can be
described as:

\smallskip
$N^*_A = \{ (\bigvee_{k \in{\mathbb N}} \xi^k_i({\overline a})) ||
  C_i({\overline a}), i = 1, {\dots}, n+k,  {\overline a} \text{ is a
    sequence of elements in } \A \}$ 

\smallskip
\noindent 
and models for $\xi_i =  \bigvee_{k \in{\mathbb N}} \xi^k_i$ 
can be built in a similar way to the way the interpretations for 
$\mu_i$ in the minimal  model for $CH_N$ are built.

\smallskip
\noindent To prove (2) note that the following are
equivalent: 
\begin{quote}
\begin{enumerate}
\item[(i)] $\A \models \exists
  P ~ N$; 
\item[(ii)] There exists a $\Pi'$-structure $\B$ with $\B \models N$
  and $\B_{|\Pi} = \A$; 
\item[(iii)] $\A \models N^*_0$;  
\item[(iv)] $\A^A \models (N^*_0)_A = (N^*_A)_0$; 
\item[(v)] $\A^A \models \neg \bigvee_{k \in {\mathbb N}}
  \xi_j({\overline a})$ for every sequence ${\overline a}$ of elements
  in $\A$; 
\item[(vi)] $\A^A \models \neg \mu^{\A}_j({\overline a})$  
for every sequence ${\overline a}$ of elements  in $\A$; 
\item[(vii)] $\A^{\mu} \models \forall {\overline x} ~ \neg
  \mu_j({\overline x})$; 
\item[(viii)] $CH_N \cup \{ \neg \mu_j({\overline  x}) \}$ is satisfiable
w.r.t.\ $\T_{\A}$; 
\end{enumerate}
\end{quote}
where $N^*_0$ ($N^*_A)_0$ is the set of all clauses in $N^*$ ($N^*_A$)
which do not contain $P$.

\noindent (i) and (ii) are equivalent by definition; (ii) and (iii) by
Theorem~\ref{thm:model-constr}; (iii) and (iv) by Lemma~\ref{lemma:instances}; 
(iv) and (v) by the fact that the 
conjunction of all clauses of $(N^*_A)_0$ can be succinctly
represented by taking a possibly infinite disjunction in the
constraint in front of $\perp$; (v) and (vi) are equivalent due to
(1); (vi) and (vii) by definition.

\noindent  (vii) $\Rightarrow$ (viii):
By assumption (vii), $\A^\mu \models \forall {\overline x} ~ \neg
  \mu_j({\overline x})$.
Since  $(\mu^{\A}_i({\overline x}))_i$ are the
interpretations of $\mu_i$ in the least model for $CH_N$ w.r.t.\
$\T_A$ it follows that 
$\A^\mu \models CH_N \cup \{ \forall {\overline x} ~ \neg
  \mu_j({\overline x}) \}$, hence $CH_N \cup \{ \forall {\overline x} ~ \neg
  \mu_j({\overline x}) \}$ is satisfiable w.r.t.\ $\T_{\A}$.

\noindent  (viii) $\Rightarrow$ (vii): Assume now that 
$CH_N \cup \{ \forall {\overline x} ~ \neg
  \mu_j({\overline x}) \}$ is satisfiable w.r.t.\ 
  $\T_{\A}$, i.e.\ there exists an expansion $\B$ of $\A$ with 
interpretations for the predicates $\mu_i$ such that 
$\B \models CH_N \cup \{ \forall {\overline x} ~ \neg
  \mu_j({\overline x}) \}$. 
Let $M^{\A}$ be the least model of
$CH_N \cup \{ \neg  \mu_j({\overline x}) \}$.  
It can be constructed with the canonical construction explained
before, by considering the set 
$(CH_N \cup \{ \neg \mu_j({\overline  x}) \})_A$ of instances of 
clauses in $CH_N \cup \{ \neg  \mu_j({\overline x}) \}$ with constants
in $\A$ and marking $\mu_i({\overline a})$ as true if we have a rule
${\sf body}(x) \rightarrow \mu_i(x)$. 
Note that $\{ \neg\mu_j({\overline  x}) \}$
does not contribute to this model building process.
This means that the least model of $CH_N \cup \{ \neg
\mu_j({\overline x}) \}$
is actually the least model of $CH_N$, hence in the least model 
of $CH_N$ the formula $\forall {\overline x} ~ \neg  \mu_j({\overline x})$ is true, which
means that $A^{\mu} \models \forall {\overline x} ~ \neg  \mu_j({\overline x})$. 
\QED

\medskip
\noindent 
If $\T$ has only one (canonical) model and is supported by  $\mu Z$ \cite{muz},
we can use $\mu Z$ for checking whether $N$ is satisfiable\footnote{If the set $N$  of constrained $P$-clauses (hence the
set of constrained Horn clauses $CH_N$) contains at least one parameter 
then $\mu Z$ often returns ``unknown''. 
In addition, 
if $\mu Z$ can prove satisfiability of $CH_N
\cup \{ \neg \mu_j({\overline x}) \}$ for a non-parametric problem, 
the model it returns is not guaranteed to be minimal in
general, and cannot be used for representing the saturated set of
clauses. By Theorem~\ref{thm:min-model} (2), satisfiability of $CH_N
\cup \{ \neg \mu_j({\overline x}) \}$ is 
sufficient for proving the satisfiability of $N$ in this case.}. 

\begin{ex}
{\em 
Consider the set $N$ consisting of the following constrained $P$-clauses:  

\smallskip
\noindent $(1)~ x = y || P(x, y), ~~ 
(2)~ y = x+1 || P(y, z) \rightarrow P(x, z), ~~ (3)~  n(x, y) || \neg P(x, y)$

\smallskip
\noindent over the theory of integers without
multiplication with model $\mathbb Z$.   
Saturating $N$ without any simplification strategy yields
the infinite set $N^*$ consisting of: 

%\medskip
\noindent {\small 
$\begin{array}{llllll} 
(1_k) & \displaystyle{\bigwedge_{i=1}^k}x_i {=} x_{i-1}{+}1 ~||~
P(x_0, x_k) & & (2_k) & \displaystyle{\bigwedge_{i=1}^k}x_i {=}
x_{i-1}{+}1 ~||~ P(x_k, z)
\rightarrow P(x_0, z) & \\
% \quad k \in {\mathbb N}, k \geq 0\\
(3_k) & n(x_0, y) \wedge \displaystyle{\bigwedge_{i=1}^k}x_i {=}
x_{i-1}{+}1  ~||~ \neg P(x_k, y) & & (4_{k}) &  
\displaystyle{\bigwedge_{i=1}^k}x_i {=} x_{i-1}{+}1 \wedge n(x_0, x_k)
~||\perp, k \in {\mathbb N}
\end{array}$
}

\smallskip
\noindent (i) We first show how Theorem \ref{thm:min-model} can be used
in this case. 
Let $N_P = \{
C_1, C_2, C_3 \}$, where $C_1 = P(x_1, y_1), C_2 = P(y_2, z_1) \rightarrow
P(x_1, z_1),$ and $C_3 = \neg P(x_3, y_3)$.
We can saturate $N_P$ as follows: From $C_1$ and $C_3$
we can derive $C_4 = \perp$; from $C_1$ and $C_2$ we can derive a clause of 
type $C_1$, from $C_2$ and $C_2$ a clause of type $C_2$ and from $C_2$
and $C_3$ a clause of type $C_3$. We obtain $N_P^* =
\{ C_1, C_2, C_3, C_4 \}$. By Theorem~\ref{thm:min-model}, the
saturation of $N$ is $N^*$: 

\smallskip
\noindent $\{ \mu_1(x,
y) ||P(x, y),\, \mu_2(x, y, z) || P(y, z) {\rightarrow} P(x, z),\,
\mu_3(x, y) || \neg P(x, y),\, \mu_4(x, y) ||\perp \}$, 

\smallskip
\noindent where $\mu_1,
\mu_2, \mu_3, \mu_4$ are given by the minimal model ${\sf M}$ of 
$CH_N$: 

\medskip
\noindent $\begin{array}{r@{}c@{}l} 
CH_N & = \{ & x = y \rightarrow \mu_1(x, y), \quad y = x + 1
\rightarrow \mu_2(x, y, z), \quad  n(x, y) \rightarrow \mu_3(x, y), \\
& &\mu_1(x, y) \wedge \mu_2(u, x, y) \rightarrow \mu_1(u, y), \quad \mu_3(x, y) \wedge \mu_2(x, u, y) \rightarrow \mu_3(u, y), \\
& &\mu_2(x, y, z) \wedge \mu_2(u, x, z) \rightarrow \mu_2(u, y, z),
\quad  \mu_1(x, y) \wedge \mu_3(x, y) \rightarrow \mu_4(x, y) \}
\end{array}$ 

\medskip
\noindent $\mu Z$ cannot check whether this set of Horn constraints is satisfiable
because of the parameter $n$. 
If we replace $n(x, y)$ with $x > y$
$\mu Z$ yields the following solution: 

\smallskip
\noindent $\mu_1(x, y) = x \leq y,~ \mu_2(x, y, z) = (y > z) \lor (x < z),~
 \mu_3(x, y) = x > y,~ \mu_4 = \perp$.

\medskip
\noindent (ii) 
Alternatively, note that if we use the fact that $\exists x_1 \dots x_{k-1} ~
\bigwedge_{i=1}^k x_i {=}
x_{i-1}{+}1 \equiv_{\T} x_k = x_0 + k$ we
obtain an infinite set of clauses consisting of: 

\medskip
\noindent $\begin{array}{llllll} 
(1'_k) & y = x+k || P(x, y) & & (2'_k) & y = x + k || P(y, z)
\rightarrow P(x, z) & \\
% \quad k \in {\mathbb N}, k \geq 0\\
(3'_k) & n(x, y) \wedge z = x + k  || \neg P(z, y) & & (4'_{k}) & y = u + k \wedge n(u, y)  || \perp & 
k \in {\mathbb N}
\end{array}$

\medskip
\noindent If we regard $k$ in each clause as a universally quantified
variable (with additional condition $k \geq 0$) we obtain: 

\smallskip
\noindent $\begin{array}{ll}
N' = \{ & y = x+k \wedge k \geq 0 || P(x, y), ~ y = x + k \wedge k
\geq 0|| P(y, z), \\
& n(x, y) \wedge z = x + k \wedge k \geq 0 || \neg P(z, y), ~ y = u +
k \wedge k \geq 0 \wedge n(u, y)  || \perp \}.
\end{array}$ 

\smallskip
\noindent If $\A = ({\mathbb Z}, n_A)$,  $\A \models \exists P ~ N'$  iff 
$\A \models \forall u, y, k ~ (k \geq 0 \wedge y = u + k \rightarrow \neg n(u,
y))$. 

\smallskip
\noindent 
{\em Remark:} In linear integer arithmetic the interpretations of $(\mu_i)_{1 \leq i
  \leq 4}$ in the minimal model of $CH_N$ w.r.t.\ the model $\A =
({\mathbb Z}, n_A)$, for a fixed interpretation of $n$ (say  as
$n_A(x,y) = (x > y)$) are: 
$\mu_1(x, y) = \mu_2(x, y, z)  = \exists k (k \geq 0 \wedge y = x + k)$, 
$\mu_3(x, y) = \exists z \exists k (n(z, y) \wedge x = z + k)$ and 
$\mu_4(x, y) = \mu_1(x, y) \wedge \mu_3(x, y)$.
} \QEX
\label{ex:horn-acceleration}
\end{ex}
Example~\ref{ex:horn-acceleration}(ii) uses 
acceleration techniques, in particular 
the following result:  
\begin{thm}[\cite{Boigelot,FinkelLeroux}] 
Let $N$ be a set of constrained clauses of the form: 

\smallskip
\noindent 
$
N = \{ \phi_0({\overline x}) ~  || ~ R({\overline x}), \quad 
           \phi({\overline x}) \wedge {\overline y} = M \cdot
           {\overline x} + {\overline v} ~  || ~ R({\overline x}) \rightarrow
           R({\overline y}) \} 
$ 

\smallskip
\noindent where ${\overline x}, {\overline y}$ describe vectors of $n$
variables, ${\overline v}$ a vector of $n$ constants in ${\mathbb Z}$,
$\phi_0$ is a condition expressible in Presburger  
arithmetic and $M = (m_{i,j})_{1 \leq i,j \leq n}$ is a $n \times n$
matrix over ${\mathbb Z}$,  
and 
$\phi(x_1, \dots, x_n) = \bigwedge_{i =    1}^k (\sum_{j = 1}^n a_{ij} x_j \leq b_i)$, where $a_{ij}, b_i
    \in {\mathbb Z}$.

The interpretation of $R$ in the minimal model of
$N$ is Presburger definable if $\left< M \right> = \{ M^n \mid n \in {\mathbb N}
\}$ is finite. 
If $\phi = \top$ then the interpretation of $R$ in the minimal model of
$N$ is Presburger definable if and only if $\left< M \right> = \{ M^n \mid n \in {\mathbb N}
\}$ is finite.  
\label{acceleration}
\end{thm}
Acceleration techniques have been investigated e.g.\ 
for fragments of theories of arrays with read
and write in the presence of iterators and selectors in
\cite{GhilardiSharyghina}. 
Similar ideas are used in the superposition
calculus in \cite{FietzkeWeidenbach,Horbach}, 
and in approaches which combine superposition and induction 
\cite{Kersani-Peltier-13} or use solutions for recurrences in loop invariant generation 
\cite{Kovacs08a,Kovacs08b}. 
We plan to analyze such aspects
in future work.

\section{Checking Entailment} 
\label{class-containment}

Let $\T$ be a theory with signature $\Pi = (S, \Sigma,
{\sf Pred})$, and let 
 ${\overline P}_1 = P^1_1, \dots, P^1_{n_1}$ and ${\overline P}_2 = P^2_1,
 \dots, P^2_{n_2}$ be finite sequences of different predicate symbols with
 $P^i_j \not\in {\sf Pred}$, and 
$\Pi_{i} = (\Sigma, {\sf Pred} \cup \{ P^i_j \mid 1 \leq j \leq n_i
\})$ 
for $i = 1, 2$.

\noindent Let $F_1$ be a universal $\Pi_{1}$-formula and  $F_2$ be a
universal $\Pi_{2}$-formula. 
We analyze the problem of checking whether 
``$\exists {\overline P}_1 ~ F_1$ entails $\exists {\overline P}_2 ~ F_2$ w.r.t.\ $\T$'' holds.

\begin{ex}
{\em Such questions arise in the graph-theoretic problems
discussed in Section~\ref{networks}. 
Let $\class{A}$ be a class of graphs described by axioms ${\sf Ax}_A$
and  $\class{B}$ be a class of graphs described by axioms ${\sf
  Ax}_B$. Let $\T$ be a theory used for expressing these axioms. 
Consider the $\cdot^+$ and $\cdot^-$ transformations
described in Section~\ref{networks}. 
Then $\class{A}^+ \subseteq \class{B}^-$ (i.e.\
 for every graph $H = (V, F) \in \class{A}^+$ we have $H \in
  \class{B}^-$) if and only if  
$\exists E_A ~ ({\sf Ax}_A \wedge Tr^+(E_A, F)) \models_{\T}
  \exists E_B ~ ({\sf Ax}_B \wedge Tr^-(E_B, F))$. 
} \QEX 
\end{ex}

\noindent 
Assume that there exist $\Pi$-formulae $G_1$ and $G_2$ such that 
$G_1 \equiv_{\T} \exists {\overline P}_1 F_1$ and $G_2 \equiv_{\T}
\exists {\overline P}_2 F_2$.
Such formulae can be found either
by saturation\footnote{We
can iterate the application of $\mathit{HRes}^P_{\succ}$ for
variables $P^i_1, \dots, P^i_n$ (in this order). This corresponds to a
variant of ordered resolution which we denote by
$\mathit{HRes}^{P^i_1,\dots,P^i_n}_{\succ}$; 
if saturation terminates the
conjunction of clauses not containing $P^i_1,\dots, P^i_n$ is equivalent
to $\exists P^i_1,\dots,P^i_n~ N_{F_i}$, where $N_{F_i}$ is the clause
form of $F_i$.} 
by successively
eliminating $P_1, \dots, P_n$, or by using
acceleration techniques or other methods.
In this case, $\exists {\overline P}_1 ~ F_1 \models_{\T} \exists
{\overline P}_2 ~ F_2$ iff $G_1 \models_{\T} G_2$
(which is the case iff $G_1 \wedge \neg G_2 \models_{\T} \perp$).

\smallskip
\noindent 
The problem of checking whether $G_1 \wedge \neg G_2\models_{\T}
\perp$ is in general undecidable, even if $G_1$ and
$G_2$ are universal formulae and $\T$ is the
extension of Presburger arithmetic or real arithmetic with a new
function or predicate symbol (cf.\ 
\cite{Voigt-thesis}). 

\smallskip
\noindent If $G_1 \wedge \neg G_2$ is in a fragment of  $\T$ for which 
checking satisfiability is decidable, then we can effectively check
whether $\exists {\overline P}_1 ~ F_1 \models_{\T} \exists \overline{P}_2 ~ F_2$. 
This is obviously the case when $\T$ is a
decidable theory. We will show that a similar condition can be
obtained for local theory extensions of theories allowing quantifier
elimination if $G_1$ and $G_2$ are universal formulae and 
the extensions satisfy a certain ``flatness property''
which allows finite complete instantiation and that in both cases we
can also generate constraints on ``parameters'' under which entailment holds. 
\begin{thm}
\label{entailment-qe}
Assume that 
there exist $\Pi$-formulae $G_1$ and $G_2$ such that 
$G_1 \equiv_{\T} \exists {\overline P}_1 F_1$ and $G_2 \equiv_{\T}
\exists {\overline P}_2 F_2$.
If $\T$ is a decidable theory then 
we can effectively check whether $\exists {\overline P}_1 ~ F_1 \models_{\T}
\exists {\overline P}_2 ~ F_2$. If $\T$ has quantifier elimination and the
formulae $F_1, F_2$ contain parametric constants, we can use
quantifier elimination in $\T$ to derive conditions on these
parameters under which $\exists {\overline P}_1 \, F_1 \models_{\T}
\exists {\overline P}_2 \, F_2$.
\end{thm}

\begin{thm}
Assume that 
there exist {\em universal} $\Pi$-formulae $G_1$ and $G_2$ such that 
$G_1 \equiv_{\T} \exists {\overline P}_1 F_1$ and $G_2 \equiv_{\T}
\exists {\overline P}_2 F_2$, and that $\T = \T_0 \cup \K$, where 
$\T_0$ is a decidable theory with signature $\Pi_0 = (S_0,
  \Sigma_0, {\sf Pred}_0)$ where $S_0$ is a set of interpreted sorts
  and $\K$ is a set of
  (universally quantified) clauses over $\Pi = (S_0 \cup S_1,
  \Sigma_0 \cup \Sigma_1, {\sf Pred}_0 \cup {\sf Pred}_1)$, where 
(i) $S_1$ is a new set of uninterpreted sorts, 
(ii) $\Sigma_1, {\sf Pred}_1$ are sets of new function,
resp.\ predicate symbols which have only arguments of uninterpreted
sort $\in S_1$, and all function symbols in
   $\Sigma_1$ have interpreted output sort $\in S_0$.
Assume, in addition, that all variables
and constants of sort $\in S_1$ in $\K, G_1$ and $\neg G_2$ 
occur below function symbols in $\Sigma_1$. Then: 
 \begin{itemize}
\item[(1)] We can use the decision procedure for $\T_0$ to 
   effectively check whether $G_1 \wedge \neg G_2 \models_{\T} \perp$
   (hence if $\exists {\overline P}_1 ~ F_1 \models_{\T}
  \exists {\overline P}_2 ~ F_2$). 
\item[(2)] If $\T_0$ allows quantifier elimination and the
  formulae $F_1, F_2$ (hence also $G_1, G_2$) contain parametric 
  constants and functions, we can
  use Algorithm~1 for obtaining constraints on the parameters under
  which $\exists {\overline P}_1 ~ F_1 \models_{\T}
  \exists {\overline P}_2 ~ F_2$.
\end{itemize}
\label{thm:entailment-decidable}
\end{thm}
{\em Proof:} 
Let $C$ be the set of constants
of uninterpreted sort $s \in S_1$ occurring in $\K, G_1$ and $\neg
G_2$. 
Note that $G_1 \wedge \neg G_2$ is satisfiable w.r.t.\ $\T = \T_0 \cup
\K$ iff $(\K \wedge G_1)^{[C]} \wedge \neg G_2$ is satisfiable, where 
$(\K \wedge G_1)^{[C]}$ is the set of all instances of $\K \wedge G_1$
in which the variables of sort $s \in S_1$ are replaced with constants
of sort $s$ in $C$. 
(1) The hierarchical reasoning method in
Theorem~\ref{lemma-rel-transl} allows us to
reduce testing whether $G_1 \wedge \neg G_2 \models_{\T} \perp$ to 
a satisfiability test w.r.t.\ $\T_0$. 
(2)  If $\T_0$ allows QE we can use Theorem~\ref{thm-alg-1}. 
 \QED

\subsection{Application: Checking class inclusion}

\noindent 
We illustrate how Theorem~\ref{thm:entailment-decidable} can be used for checking
one of the class inclusions mentioned in Section~\ref{networks}. 

\begin{ex}
\label{qudg}
{\em Let $\class{QUDG}(r) = ( \class{MinDG}(r) \cap \class{MaxDG}(1) )^-$, 
be axiomatized by ${\sf MinDG}(r)
\wedge {\sf MinDG}(1) \wedge {\sf Tr^-}(E, F)$, where: 

\medskip
\noindent 
$\begin{array}{llll}
{\sf MinDG}(r): & \forall x, y & \pi^i(x, y, r) \rightarrow E(x, y) &
\quad \!\!\!\!\!\!\!\!\!\!\!\!\!\!\!\!\!\!\!\!\! \!\!\!\!\!\!\!\!\!\!\! \text{
  where } \pi^i(x, y, r) = x \neq y \land d(x,y) \leq r(x) \\
{\sf MaxDG}(1): & \forall x, y & \pi^e(x, y) \rightarrow \lnot E(x, y) &
\quad \!\!\!\!\!\!\!\!\!\!\!\!\!\!\!\!\!\!\!\!\! \!\!\!\!\!\!\!\!\!\!\! \text{ where } \pi^e(x, y) = d(x,y) > 1\\
{\sf Tr^-}(E, F): & \forall x, y & (F(x, y)  \leftrightarrow E(x, y)
\land E(y, x)) & .
\end{array}$

\medskip
\noindent 
We want to check whether $\class{A}(r) \subseteq \class{B}(r)$, where 
$\class{A}(r) = \class{QUDG}(r)$ and 
$\class{B}(r) = ( \class{MinDG}(r) \cap \class{MaxDG}(1) )^+$ is
described by ${\sf MinDG}(r)
\wedge {\sf MinDG}(1) \wedge {\sf Tr^+}(E, F)$.

\medskip
\noindent We  obtain the  
axiomatization $G_1$ by eliminating $E$ from 
$$\exists E ({\sf MinDG}(r)
\wedge {\sf MinDG}(1) \wedge {\sf Tr^-}(E, F))$$ 
and  the axiomatization 
$G_2$ by eliminating $E$ from 
 $$\exists E ({\sf MinDG}(r)
\wedge {\sf MinDG}(1) \wedge {\sf Tr^+}(E, F)).$$

\medskip
{\small \noindent $\begin{array}{lll} 
& G_1 & \\
\hline 
\forall x, y ~ &   \pi^i(x,y,r) \land \pi^e(x,y) & \rightarrow ~~ \perp\\
\forall x, y &  \pi^i(x,y) \land \pi^i(y,x) & \rightarrow ~~ F(y,x) \\
\forall x, y &   \pi^e(x,y) & \rightarrow \neg F(x,y) \\
\forall x, y &   \pi^e(x,y) & \rightarrow \neg F(y,x) \\
\forall x, y &   F(x,y) & \rightarrow ~~ F(y,x) \\
& & 
\end{array}$ \quad 
$\begin{array}{lll} 
& G_2 & \\
\hline 
\forall x, y ~ &   \pi^i(x,y,r) \land \pi^e(x,y) & \rightarrow ~~ \perp\\
\forall x, y &  \pi^e(x,y) \land \pi^e(y,x) & \rightarrow \neg F(y,x) \\
\forall x, y &   \pi^i(x,y) & \rightarrow ~~ F(x,y) \\
\forall x, y &   \pi^i(x,y) & \rightarrow ~~ F(y,x) \\
\forall x, y &   F(x,y) & \rightarrow ~~ F(y,x) \\
\forall x & \pi^e(x, x) & \rightarrow \neg F(x, x)
\end{array}$
} 

\smallskip
\noindent We check whether $G_1 \models_{\T} G_2$, i.e.\ 
whether $G_1 \wedge \neg G_2$ is unsatisfiable w.r.t.\ $\T$, where 
$\neg G_2$ is the disjunction of the following
ground formulae (we ignore the negation of the first clause obviously
implied by $G_1$): 

\smallskip
\noindent 
{\small $\begin{array}{llllllll} 
~ (g_1) & \pi^e(a, b) \wedge \pi^e(b, a) \wedge F(b, a) & \quad & (g_2) &
\pi^e(a, a) \wedge F(a, a) & \quad & (g_3) & F(a, b) \wedge \neg F(b, a) \\
~ (g_4) & \pi^i(a, b) \wedge \neg F(a, b) & &  (g_5) & \pi^i(a, b)
\wedge \neg F(b, a) \\
\end{array}$
}

\smallskip
\noindent 
By Theorem~\ref{thm:entailment-decidable} (2), we can consider  
the set of all instances of $G_1$ in which the variables of sort ${\sf
  p}$ are replaced with 
the constants $a, b$, then use a
method for checking ground satisfiability of $G_1[T] \wedge g_i$ 
w.r.t.\  $\T^u_d$ ($d$ is uninterpreted), $\T^p_d$ ($d$ is positive),
$\T^s_d$ ($d$ is symmetric) and $\T^m_d$ ($d$ is a metric). For this,
we use H-PILoT \cite{hpilot} in which we enforce the right
instantiation by adding relevant instances to the query.  
This allows us to check that $G_1[T] \wedge g_i$ is unsatisfiable for 
$i \in \{ 1, 2, 3 \}$, but satisfiable for $i \in \{ 4,5 \}$ 
(this is so for all four theories). 

For cases 4 and 5 we use an implementation of Algorithm 1, {\sf sehpilot}
 to derive conditions on parameters under which $G_1[T] \wedge g_i$ 
is unsatisfiable.
We give here two examples: 

\medskip
\noindent (1) We consider $d$ and $r$ to be
  parameters,  i.e. we eliminate only $F$ from $G_1[T] \wedge g_i$. 
For $\T^m_d$ we get the condition 

\smallskip
$C^{d,r} = \forall x, y (x \neq y \wedge
d(x, y) \leq 1 \wedge
d(x, y) \leq r(x) \rightarrow d(y, x) \leq r(y)).$

\medskip
\noindent (2) We consider only $r$ to be a parameter, i.e.\  
we eliminate the symbols $F$ and $d$. 
For $\T^s_d$ 
we obtain the condition 

\smallskip
$C^{r} =  \forall x, y (r(y) < 1  \wedge
x \neq y  \rightarrow r(y) \geq r(x)).$

\smallskip
\noindent This condition holds e.g.\ if
$r(x)=r(y)$  for all $x,y$, i.e. if $r$ is a constant function. Adding this as an additional condition we get unsatisfiability of $G_1[T] \wedge g_i$ with $i \in \{ 4,5 \}$ for $\T^m_d$ and $\T^s_d$, but not for $\T^u_d$ and $\T^p_d$.

\

\smallskip
\noindent 
\textbf{Checking the other inclusion}
We now check whether $\class{B}(r) \subseteq \class{A}(r)$, where 
$\class{A}(r) = \class{QUDG}(r)$ and 
$\class{B}(r) = ( \class{MinDG}(r) \cap \class{MaxDG}(1) )^+$.
We  have the axiomatizations $G_1$, $G_2$ for the two classes.

\smallskip
\noindent We check whether $G_2 \models_{\T} G_1$, i.e.\ 
whether $G_2 \wedge \neg G_1$ is unsatisfiable w.r.t.\ $\T$, where 
$\neg G_1$ is the disjunction of the following
ground formulae (we ignore the negation of the first clause obviously
implied by $G_2$): 

\smallskip
\noindent 
{\small $\begin{array}{llllllll} 
~ (g_1) & \pi^i(a, b) \wedge \pi^i(b, a) \wedge \neg F(b, a) & \quad & (g_2) &
\pi^e(a, b) \wedge F(a, b) \\
~ (g_3) & \pi^i(a, b) \wedge \neg F(b, a) & &  (g_4) & F(a, b)
\wedge \neg F(b, a) \\
\end{array}$
}

\smallskip
\noindent 
We use H-PILoT for checking ground satisfiability of $G_2[T] \wedge g_i$ 
w.r.t.\  $\T \in \{ \T^u_d, \T^p_d, \T^s_d, \T^m_d \}$. For $T_s$ and $T_m$ we obtain unsatisfiability of $G_2[T] \wedge g_i$ for 
$i \in \{ 1, 2, 3, 4 \}$, thus we have proved that the inclusion holds for these two theories.
For $T_p$ and $T_u$ we get satisfiability for cases 2 and 3. We use Algorithm 1 to obtain conditions on parameters such that $G_2[T] \wedge g_2$ and $G_2[T] \wedge g_3$ is unsatisfiable.

\smallskip
\noindent If we consider $d$ and $r$ to be parameters, i.e. we eliminate only
$F$ from $G_1[T] \wedge g_i$ we obtain the condition 

\smallskip
$C^{d,r} = \forall x, y (d(y,x)>1 \lor d(x,y) \leq 1 \lor d(x,y) \leq
r(x) \lor x=y).$ 

\smallskip
\noindent It is easy to see that this condition holds if $d$ is symmetric.
\QEX

}

\end{ex}

\section{Tests}
\label{tests}

We tested the methods we proposed on several examples. 
We used various tools for solving the various types of symbol
elimination considered in this paper. 

\medskip
\noindent {\bf Second-order quantifier elimination.}
Since the  implementations of the hierarchical superposition calculus we are
aware of have as background theory linear arithmetic and in our 
examples we had more complex theories, we used a form of 
abstraction first: We renamed the constraints over more complex 
theories with new predicate symbols, and used SCAN 
\cite{Gabbay-Ohlbach} for second-order quantifier
elimination. 
SCAN performs second-order quantifier elimination in first-order
logic. It takes as input a formula of the form $F(P_1,\dots, P_n)$ 
containing predicate symbols $P_1, \dots, P_n$ and  
applies a clause form transformation, ordered
resolution and de-Skolemization on this formula. 
In case of termination and if de-Skolemization is possible,  it
returns a first-order formula equivalent to $\exists P_1, \dots
\exists P_n F(P_1, \dots, P_n)$, which does not contain the predicate 
symbols $P_1, \dots, P_n$.

\medskip
\noindent {\bf Satisfiability checking and property-directed symbol elimination.} 
For satisfiability checking we used H-PILoT \cite{hpilot}
(after preparing the input such that the instances that have to be
used are clear for the prover).
H-PILoT carries out a hierarchical reduction to the base theory. 
Standard SMT provers or specialized provers can
be used for testing the satisfiability of the formulae obtained after
the reduction.
H-PILoT uses eager instantiation and the hierarchical reduction, 
so provers like CVC4 \cite{cvc} or Z3 \cite{z3-2020,z3-2018} are in general faster in proving 
unsatisfiability. The advantage of using H-PILoT is that knowing 
the instances needed for a complete instantiation allows us to correctly detect
satisfiability (and generate models) in situations in which e.g.\ CVC4
returns ``unknown'', and use 
property-directed symbol elimination 
to obtain additional constraints on parameters which 
ensure unsatisfiability.

For obtaining the constraints on parameters 
we used the method described in Algorithm~\ref{algorithm-symb-elim} 
proposed in \cite{Sofronie-lmcs2018} which was implemented in 
{\sf  sehpilot} for the case in which the base theory is the theory of
real-closed fields.
{\sf sehpilot} (Symbol Elimination with H-PILoT) receives a list of 
parameters as a command line (and possibly a list of already existing 
constraints on these parameters) and 
uses H-PILoT for the
hierarchical reduction to a problem in the base theory (Step 1 in
Algorithm~\ref{algorithm-symb-elim}) and for generating a 
corresponding REDLOG file. The constants are classified as required in
Step 2 of Algorithm~\ref{algorithm-symb-elim} and the 
REDLOG file is changed accordingly such that  only those symbols that
are not a parameter or argument of a parameter
are considered to be existentially quantified. Redlog is used for
quantifier elimination (Step 3 of
Algorithm~\ref{algorithm-symb-elim}); then the constants contained 
in the obtained formula are replaced back with the terms they
represent (Step 4). Finally, the formula obtained this way is negated 
(Redlog is used for further simplifications).

\noindent The way we used these tools is illustrated on some tests in Appendix~\ref{app:examples}.

\section{Conclusions}
\label{conclusions}

In this paper, we analyzed possibilities of combining 
general second-order symbol elimination 
and property-directed symbol elimination. 
For eliminating existentially quantified
predicates from universal first-order formulae 
we used a constrained resolution calculus (obtained from specializing the hierarchical
superposition calculus).  We analyzed situations in which saturation
terminates and two possibilities of obtaining finite representations 
also in cases in which saturation might not terminate: (i) Using an
encoding of the constraints of the saturated set of clauses 
as smallest fixpoints of certain families
of constrained Horn clauses and (ii) using acceleration.  
For checking the satisfiability of families of constrained Horn
clauses we used the fixpoint package of
Z3 \cite{muz}. 

If the saturation terminates, or the infinite saturated
set of clauses has a finite representation, we can use the obtained
set of clauses for checking entailment. We proved a $\Psi$-locality 
property for a class of formulae; this allowed us to use the prover
H-PILoT  (after preparing the input such that the instances that have to be
used are clear) for analyzing the satisfiability of formulae w.r.t. models in a theory $\T$
and for checking entailment between formulae.
Property-based symbol elimination proved useful for obtaining  
(weakest) constraints $\Gamma$ on ``parameters'' used in the description of the
theory $\T$ such that satisfiability or entailment is guaranteed 
in models satisfying $\Gamma$.
 
In future work we would like to find possibilities of identifying
situations in which second-order quantifier elimination using
resolution terminates and study possibilities of using (and
generalizing) methods based on constrained Horn clauses or 
acceleration for obtaining finite representations of potentially
infinite clause sets. 
We would also like to analyze possibilities of checking
entailment when the second-order
quantifier elimination method returns a fixpoint and not a formula. 
(The main obstacle when working on this problem was that
$\mu Z$ returns ``unknown'' in the presence of parameters.)

\

\noindent {\bf Acknowledgments:} We thank Hannes Frey and Lucas
B{\"o}ltz for the numerous discussions we had on the problems in
wireless networks discussed in Section~\ref{networks}, 
Renate Schmidt for maintaining a website where one can run SCAN online 
and for sending us the executables and instructions for running them. 
We thank the reviewers for their helpful comments.

\newpage
\appendix

\section{Tests}
\label{app:examples}

We here present some of the tests we made for the examples in the
paper. We show how we used the tools on these 
examples including the corresponding input and output files.

\subsection{Tests for Example \ref{ex-sat2}}
We used H-PILoT to check that $(c1)$, $(c2)$ and $(c3)$ from Example
$\ref{ex-sat1}$ are valid w.r.t. $\T = \T_d^m \cup {\sf Free}(r_1, r_2)$. 
For this we show, one after the other, that the negation of each of
these formulae is unsatisfiable w.r.t.\ $\T$.
We start with $(c1)$. 
In the input file for H-PILoT we have the axioms for a metric specified 
under {\bf Clauses} and the negation of $(c1)$ is the {\bf Query}.

{\small 
\begin{lstlisting}[frame=single]
Base_functions:={(+,2), (-,2), (*,2)}
Extension_functions:={(r1, 1, 1), (r2, 1, 1), (d, 2, 1)}
Relations:={(<=, 2), (<, 2), (>=, 2), (>, 2)}

Clauses :=  (FORALL x,y). d(x,y) = _0 --> x = y;
            (FORALL x,y). x = y --> d(x,y) = _0;
            (FORALL x,y). d(x,y) = d(y,x);
            (FORALL x,y,z). d(x,y) <= d(x,z) + d(z,y);

Query :=    NOT(u = v); 
            d(u,v) <= r1(u); 
            NOT(u = w); 
            d(u,w) <= d(u,v); 
            OR(u = w, d(u,w) > r1(u));
\end{lstlisting}
}

\noindent H-PILoT performs the hierarchical reduction described in 
Theorem~\ref{lemma:instances}, then hands the reduced problem over to
a prover to check for satisfiability. 
We here used Z3, which is also the default prover used by H-PILoT. 
We obtain the following output from H-PILoT:

{\small \begin{lstlisting}[frame=single]
Reduced problem written to 
example4-check_redundancy_1.smt.
unsat
H-PILoT spent                0.013423s on the problem.
The prover needed            0.012652s for the problem.
Total running time:          0.026075s.
\end{lstlisting}
}

\noindent The answer is ``unsat'' (unsatisfiable), so we have 
proved that $(c1)$ is valid w.r.t. $\T = \T_d^m \cup {\sf Free}(r_1,
r_2)$. 
In the same way we can prove validity of $(c2)$ and $(c3)$  w.r.t. 
$\T_d^m\cup {\sf Free}(r_1,r_2)$, 
and also the validity of the three formulae w.r.t. $\T_d^u\cup {\sf  Free}(r_1,r_2)$, 
$\T_d^p\cup {\sf Free}(r_1,r_2)$ and $\T_d^n\cup {\sf Free}(r_1,r_2)$.

\

\noindent {\bf Remark.} The encoding in H-PILoT presented above did
not use two different sorts ${\sf p}$ and ${\sf num}$ as described in
the theoretical considerations. Since in our case the sort ${\sf p}$
can be considered to be uninterpreted and there are no function
symbols of arity ${\sf p}^k \rightarrow {\sf p}$, the following holds:
Let $G$ be a flat ground formula over a signature containing a binary
function $d$ and a unary function $r_1$ with the property that the only 
constraints on the constants used as arguments for $d$ and $r_1$ 
are equalities and disequalities. Then the following are equivalent:
\begin{itemize}
\item[(1)] $G$ is satisfiable w.r.t.\ the extension
  of ${\mathbb R}$ with a binary function $d$ satisfying the metric
  axioms and a unary free function symbol $r_1$. 
\item[(2)] $G$ is satisfiable w.r.t.\ the extension of the 
two-sorted theory $\T^m_d$ with the free function $r_1$.
\end{itemize}
Indeed, from every model $\A = ({\mathbb R}, d_A : {\mathbb R}^2
\rightarrow {\mathbb R}, 
{r_1}_A : {\mathbb R} \rightarrow {\mathbb R})$ 
of $G$ w.r.t.\ the extension  of ${\mathbb R}$ with a binary function $d$ satisfying the metric
  axioms and a unary free function symbol $r_1$, 
we can define a model 
$$\B = (B_{\sf p}, {\mathbb R}, d_B: B_{\sf p}^2
\rightarrow {\mathbb R}, {r_1}_B : B_{\sf p} \rightarrow {\mathbb R})$$ 
of $\T^m_d \cup {\sf Free}(r_1)$ as follows: 
% with universe of sort {\sf p} 
% \marginpar{DP: {\sf p} $\B_{\sf p}$ looks confusing. What is $\B_{\sf p}$ here?}
\begin{itemize}
\item Take as $\B_{\sf p}$ an isomorphic copy (via isomorphism $i$) 
of  the set  \\
$\{ c_{\A} \mid c \text{ constant of sort  {\sf p}  occurring in } G \}$, 
\item Define $d_B, {r_1}_B$ as follows:
\begin{itemize}
\item $d_{\B}(i(c_{\A}), i(d_{\A})) := d_{\A}(c_{\A}, d_{\A})$;
\item ${r_1}_{\B}(i(c_{\A})) := {r_1}_{\A}(c_{\A})$.
\end{itemize}
\end{itemize}
The converse implication is analogous, with the only difference that
we first construct a partial algebra 
$\A = ({\mathbb R}, d_A : {\mathbb R}^2
\rightarrow {\mathbb R}, 
{r_1}_A : {\mathbb R} \rightarrow {\mathbb R})$  by considering an 
injective map from the support of sort ${\sf p}$ in ${\mathbb R}$
and then we use the locality property of $\T^m_d \cup {\sf Free}(r_1)$
to prove the existence of a total model with support ${\mathbb R}$.

\subsection{Tests for Example \ref{ex:sat}}
We use sehpilot (an implementation of Algorithm 1) to derive  constraints $\Gamma$ on the parameters
$r_1, r_2$ such that

\medskip
$\pi^i(u,v) \wedge \pi^e(u,v) = (u \neq v \land d(u,v) \leq r_1(u))
\wedge (d(u, v) > r_2(u))$

\medskip
\noindent 
is unsatisfiable (we consider the case in which $d$ is an uninterpreted function).
Since sehpilot first uses H-PILoT for the hierarchical reduction (and afterwards Redlog for quantifier elimination), the input file is in H-PILoT syntax:

{\small 
\begin{lstlisting}[frame=single]
Base_functions:={(+,2), (-,2), (*,2)}
Extension_functions:={(r1, 1, 1), (r2, 1, 1), (d, 2, 1)}
Relations:={(<=, 2), (<, 2), (>=, 2), (>, 2)}
	
Query :=	NOT(u = v);
		d(u,v) <= r1(u);
		d(u,v) > r2(u);
\end{lstlisting}
}

\noindent 
Note that when using sehpilot the user has to specify which symbols
have to be eliminated. 

Assume that $r_1$ and $r_2$ are parameters (and thus should not be
eliminated).
Since the variable $u$ occurs as an argument of $r_1$ and $r_2$ (which are
parameters), $u$ should not be eliminated. 
We have to eliminate the remaining symbols, i.e. $v$ and $d$. 
We obtain the following output (in verbose mode) from sehpilot:

{\scriptsize 
\begin{lstlisting}[frame=single]
[2021-02-12 14:06:07,115 | INFO] convert prefix notation
{'e_3': 'r2(u)', 'e_2': 'r1(u)', 'e_1': 'd(u, v)'}
to
{'e_3': 'r2(u)', 'e_2': 'r1(u)', 'e_1': 'd(u, v)'}

[2021-02-12 14:06:07,115 | INFO] assignments of new variables
e_3 = r2(u)
e_2 = r1(u)
e_1 = d(u, v)

[2021-02-12 14:06:07,115 | INFO] reduced assignments
e_3 = r2(u)
e_2 = r1(u)
e_1 = d(u, v)

[2021-02-12 14:06:07,115 | INFO] variables that will be eliminated 
by REDLOG:
e_1, d, v

[2021-02-12 14:06:07,116 | INFO] extension functions declared in loc-file
[('r1', 1, 1), ('r2', 1, 1), ('d', 2, 1)]

[2021-02-12 14:06:07,116 | INFO] universal arguments
u, u

[2021-02-12 14:06:07,116 | INFO] add switches:
off nat;

[2021-02-12 14:06:07,117 | INFO] saved REDLOG file with 
new variable declaration
/home/dpeuter/Work/FroCoS-2021/example4.dat

[2021-02-12 14:06:07,315 | INFO] REDLOG query after 
variable elimination:
all(u, e_2 - e_3 > 0)

[2021-02-12 14:06:07,315 | INFO] REDLOG command:
run_redlog_computation := not (e_2 - e_3 > 0);

[2021-02-12 14:06:07,508 | INFO] REDLOG command:
run_redlog_computation := rlnnf (not(e_2 - e_3 > 0));

[2021-02-12 14:06:07,707 | INFO] reduce constants {} of
e_2 - e_3 <= 0
to
e_2 - e_3 <= 0

[2021-02-12 14:06:07,707 | INFO] REDLOG command:
run_redlog_computation := rlsimpl (e_2 - e_3 <= 0);

Constraints (H-PILoT syntax, reduced):
(FORALL u). r1(u) - r2(u) <= _0
\end{lstlisting}
} 

\noindent 
The generated constraint $\Gamma=\forall u (r_1(u) \leq r_2(u))$ is
exactly the constraint we obtained by applying Steps 1-5 of Algorithm~1 by hand 
in Example \ref{ex:sat}. 

\smallskip
\noindent We used verbose mode for the output of sehpilot such that more details are displayed in the output file. This way one can follow easily the different steps. One can for example see which new constants are introduced in the hierarchical reduction ($e_1$, $e_2$ and $e_3$) and which terms they represent. The output also shows the result obtained directly after the elimination ($e_2 > e_3$), the negation of this result ($e_2 \leq e_3)$, and finally the universally quantified formula with the constants replaced back with the corresponding terms ($\forall u (r_1(u) \leq r_2(u))$).

\

\noindent {\bf Remark:} The current implementation of {\sf sehpilot}
assumes that the problems are expressed in a local extension of the
theory of real closed fields and a reduction to quantifier elimination
in the theory of real-closed fields is performed. 
For the examples we considered this does not lead to loss of
generality because the 
constraints on constants of sort ${\sf p}$  are only equalities and disequalities. 
If variables initially of sort ${\sf p}$ are
  eliminated, they do not occur below any parameter. 
Such variables occur separately from the variables of 
original sort ${\sf num}$ in the quantifier elimination problem.

\noindent This means that the quantifier elimination problem is of the form 
$$\exists x_1, \dots, x_n \exists y_1, \dots, y_m C_{\sf p}(x_1,
\dots, x_n) \wedge C_{\sf num}(y_1, \dots, y_m)$$
where $x_1, \dots, x_n$ are variables of sort ${\sf p}$ and 
$y_1, \dots, y_m$ are variables of sort ${\sf num}$, 
which is equivalent to: 
$$ \exists x_1, \dots, x_n C_{\sf p}(x_1,
\dots, x_n) ~~\wedge~~  \exists y_1, \dots, y_m  C_{\sf num}(y_1,
\dots, y_m).$$
Quantifier elimination in the theory of real-closed fields can be used
for the formula $\exists y_1, \dots, y_m  C_{\sf num}(y_1,
\dots, y_m)$. 

\smallskip
\noindent If we consider theories whose models of sort ${\sf p}$ contain
infinitely many elements, then -- since the constraint 
$C_{\sf p}$ contains only equalities and disequalities -- 
the method for quantifier elimination in the theory of infinite sets 
can be simulated by the method for quantifier elimination in real
closed fields. This is the reason why for this type of problems we can 
use quantifier elimination in the theory of real closed fields without 
problems.

\subsection{Tests for Example \ref{qudg}}
In order to check whether the class containment 
$$\class{QUDG}(r) = ( \class{MinDG}(r) \cap \class{MaxDG}(1) )^-
\subseteq (\class{MinDG}(r) \cap \class{MaxDG}(1) )^+$$ 
holds we have to check whether $G_1 \land g_i$ is unsatisfiable for
all $i \in {1,2,3,4,5}$ 
(where $G_1$ is the axiomatization for $\class{QUDG}(r)$ and the 
$g_i$ are the ground formulae obtained from the negation of $G_2$, 
the axiomatization of the other class; cf. Example \ref{qudg}).

We assume that $d$ is a metric. Using H-PILoT we can show that 
$G_1 \land g_i$ is unsatisfiable w.r.t. $\T_d^m$ for $i \in
\{1,2,3\}$. 
We here only show the test for the case $G_1 \land g_4$ in detail (the
case $G_1 \land g_5$ 
is similar and yields the same results). 

\smallskip
\noindent 
We check satisfiability of $G_1 \land g_4$ w.r.t. $\T_d^m$ using H-PILoT. We have the following input file:

{\footnotesize \begin{lstlisting}[frame=single]
Base_functions:={(+,2), (-,2), (*,2)}
Extension_functions:={(r, 1, 1), (d, 2, 1), (F, 2, 1)}
Relations:={(<=, 2), (<, 2), (>=, 2)}
	
Clauses :=

% axioms for G1
(FORALL x,y). d(x,y) <= r(x), d(x,y) > _1 
		--> x = y, _0 = _1;
(FORALL x,y). d(x,y) <= r(x), d(y,x) <= r(y) 
		--> x = y, F(y,x) = _1;
(FORALL x,y). d(x,y) > _1 --> F(x,y) = _0;
(FORALL x,y). d(x,y) > _1 --> F(y,x) = _0;
(FORALL x,y). F(x,y) = _1 --> F(y,x) = _1;

% axioms for d being a metric
(FORALL x,y). d(x,y) >= _0;
(FORALL x,y). d(x,y) = _0 --> x = y;
(FORALL x,y). x = y --> d(x,y) = _0;
(FORALL x,y). d(x,y) = d(y,x);
(FORALL x,y,z). d(x,y) <= d(x,z) + d(z,y);

% F is either 0 or 1
(FORALL x,y). --> F(x,y) = _0, F(x,y) = _1;
			
Query :=	

% g4 of (not G2)
NOT(a = b);
d(a,b) <= r(a);
F(a,b) = _0;

% needed for instantiation
F(a,b) = F(a,b); 
F(b,a) = F(b,a); 
F(a,a) = F(a,a); 
F(b,b) = F(b,b);

d(a,b) = d(a,b); 
d(b,a) = d(b,a); 
d(a,a) = d(a,a); 
d(b,b) = d(b,b);

r(a) = r(a); 
r(b) = r(b);}
\end{lstlisting}
}

\noindent 
Note that the trivial equalities at the end of the file are used to
ensure that H-PILoT computes sufficiently many instances. 
We obtain the following output from H-PILoT:

{\small 
\begin{lstlisting}[frame=single]
Reduced problem written to example8-Tm-4.smt.
Unknown. Prover says 'sat' but this can only be trusted 
for local extensions and this problem is not known to be 
local.
H-PILoT spent                0.208608s on the problem.
The prover needed            0.01677 s for the problem.
Total running time:          0.225378s.
\end{lstlisting}
}

\noindent Since we know that $\T^m_d \cup {\sf Free}(r_1, r_2)$ is a
$\Psi$-local theory extension and we ensured that H-PILoT computes
sufficiently many instances, we know that $G_1 \wedge g_4$ is satisfiable.
This means that the class inclusion does not hold in general. 
We use {\sf sehpilot} to derive (weakest) conditions $\Gamma$ on parameters such 
that unsatisfiability of $G_1 \land \Gamma \land g_4$ is guaranteed. 

\medskip
\noindent 
We first consider $r$ to be the only parameter, i.e. we tell sehpilot to eliminate $F$ and $d$ ($a$ and $b$ appear as arguments of parameter $r$ and are therefore not eliminated). The input file is the same file that was used for checking satisfiability with H-PILoT. We get the following output (using verbose mode) from sehpilot:

{\scriptsize 
\begin{lstlisting}[frame=single]
[2021-02-20 12:55:56,356 | INFO] convert prefix notation
{'e_10': 'r(b)', 'e_9': 'r(a)', 'e_8': 'd(b, b)', 
'e_7': 'd(b, a)', 'e_6': 'd(a, b)', 'e_5': 'd(a, a)', 
'e_4': 'F(b, b)', 'e_3': 'F(b, a)', 'e_2': 'F(a, b)', 
'e_1': 'F(a, a)'}
to
{'e_10': 'r(b)', 'e_9': 'r(a)', 'e_8': 'd(b, b)', 
'e_7': 'd(b, a)', 'e_6': 'd(a, b)', 'e_5': 'd(a, a)', 
'e_4': 'F(b, b)', 'e_3': 'F(b, a)', 'e_2': 'F(a, b)', 
'e_1': 'F(a, a)'}

[2021-02-20 12:55:56,356 | INFO] assignments of new variables
e_10 = r(b)
e_9 = r(a)
e_8 = d(b, b)
e_7 = d(b, a)
e_6 = d(a, b)
e_5 = d(a, a)
e_4 = F(b, b)
e_3 = F(b, a)
e_2 = F(a, b)
e_1 = F(a, a)

[2021-02-20 12:55:56,356 | INFO] reduced assignments
e_10 = r(b)
e_9 = r(a)
e_8 = d(b, b)
e_7 = d(b, a)
e_6 = d(a, b)
e_5 = d(a, a)
e_4 = F(b, b)
e_3 = F(b, a)
e_2 = F(a, b)
e_1 = F(a, a)

[2021-02-20 12:55:56,356 | INFO] variables that will be 
eliminated by REDLOG:
e_1, e_7, e_6, e_3, d, e_5, e_8, e_4, F, e_2

[2021-02-20 12:55:56,357 | INFO] extension functions 
declared in loc-file
[('r', 1, 1), ('d', 2, 1), ('F', 2, 1)]

[2021-02-20 12:55:56,357 | INFO] universal arguments
b, a

[2021-02-20 12:55:56,358 | INFO] add switches:
off nat;

[2021-02-20 12:55:56,358 | INFO] saved REDLOG file with 
new variable declaration
/home/dpeuter/Work/FroCoS-2021/example8/example8-Tm-4.dat

[2021-02-20 12:55:56,631 | INFO] REDLOG query after 
variable elimination:
all({a, b}, e_9 > 0 and e_9 - 1 <= 0 and e_10 - e_9 < 0 
and a - b <> 0 or e_9 - 1 
>= 0 and e_10 - 1 < 0 and a - b <> 0)

[2021-02-20 12:55:56,631 | INFO] REDLOG command:
run_redlog_computation := not (e_9 > 0 and e_9 - 1 <= 0
and e_10 - e_9 < 0 and a - b <> 0 or e_9 - 1 >= 0 
and e_10 - 1 < 0 and a - b <> 0);

[2021-02-20 12:55:56,830 | INFO] REDLOG command:
run_redlog_computation := rlnnf (not(e_9 > 0 
and e_9 - 1 <= 0 and e_10 - e_9 < 0 and a - b <> 0 
or e_9 - 1 >= 0 and e_10 - 1 < 0 and a - b <> 0));

[2021-02-20 12:55:57,039 | INFO] reduce constants {} of
(e_9 <= 0 or e_9 - 1 > 0 or e_10 - e_9 >= 0 or a - b = 0) 
and (e_9 - 1 < 0 or e_10 - 1 >= 0 or a - b = 0)
to
(e_9 <= 0 or e_9 - 1 > 0 or e_10 - e_9 >= 0 or a - b = 0) 
and (e_9 - 1 < 0 or e_10 - 1 >= 0 or a - b = 0)

[2021-02-20 12:55:57,039 | INFO] REDLOG command:
run_redlog_computation := rlsimpl ((e_9 <= 0 
or e_9 - 1 > 0 or e_10 - e_9 >= 0 or a - b = 0) 
and (e_9 - 1 < 0 or e_10 - 1 >= 0 or a - b = 0));

Constraints (H-PILoT syntax, reduced):
(FORALL a, b). AND(OR(r(a) <= _0, r(a) - _1 > _0, 
r(b) - r(a) >= _0, a - b = _0), OR(r(a) - _1 < _0, 
r(b) - _1 >= _0, a - b = _0))
\end{lstlisting}
}

\noindent 
Redlog does not simplify the results of the quantifier elimination
very well, so in many cases one obtains long formulae, which sometimes
can be simplified. In this case the constraint computed by sehpilot can be simplified to

\smallskip
$C^{r} =  \forall x, y (r(y) < 1  \wedge
x \neq y  \rightarrow r(y) \geq r(x)).$
\smallskip

\noindent We could also choose different parameters, e.g. we could assume $d$ and $r$ to be parameters and then tell sehpilot to eliminate only $F$. In this case the computed constraint will be:

{\small 
\begin{lstlisting}[frame=single]
Constraints (H-PILoT syntax, reduced):
(FORALL b, a). OR(NOT(d(b, b) = _0), d(b, a) <= _0, 
d(a, b) <= _0, d(a, b) - _1 > _0, d(a, b) - r(a) > _0, 
NOT(d(a, b) - d(b, a) = _0), NOT(d(a, a) = _0), 
r(b) - d(b, a) >= _0, a - b = _0)
\end{lstlisting}
}

\noindent This constraint can be simplified to

\smallskip
$C^{d,r} = \forall x, y (x \neq y \wedge
d(x, y) \leq 1 \wedge
d(x, y) \leq r(x) \rightarrow d(y, x) \leq r(y)).$

\section{Proof of Theorem~\ref{non-linear}}
\label{app:non-linear}

\noindent {\bf Theorem~\ref{non-linear}.}
{\em Let ${\mathcal K}$ be a set of $\Sigma$-flat clauses, with the
property that every variable occurs only once in every term. 
Let 
$\Psi$ be a term closure operator with the property that 
for every flat set of ground terms $T$, $\Psi(T)$ is flat. 

\noindent Assume that $\K$ and $\Psi$
have the property that for every flat set of ground terms $T$ and 
for every clause $C \in \K$, 
if $C$ contains terms $f(x_1, \dots, x, \dots, x_n)$ and 
$g(y_1, \dots, x, \dots, y_m)$ (where $f, g \in \Sigma$ are extension
functions and $f$ and $g$ are not necessarily different), if 
$f(t_1, \dots, t, \dots, t_n),  g(s_1, \dots, s, \dots, s_m) \in \Psi_{\K}(T)$ then 
$f(t_1, \dots, s, \dots, t_n), g(s_1, \dots, t, \dots, s_m) \in \Psi_{\K}(T)$.
\noindent Then $({\sf Emb}_{w,f}^\Psi)$ implies $({\sf Loc}_f^{\Psi})$. 
}

\

% \marginpar{DP: several lines are too long in this proof}
\noindent {\em Proof:} Assume that ${\mathcal T}_0 \cup {\mathcal K}$ is not a
$\Psi$-local extension of ${\mathcal T}_0$. Then there exists 
a set $G$ of ground clauses (with additional constants) 
such that 
${\mathcal T}_0 \cup {\mathcal K} \cup G \models \perp$ but 
${\mathcal T}_0 \cup {\mathcal K}[\Psi_{\cal K}(G)] \cup G$ 
has a weak partial model $P$ 
in which all terms in $\Psi_{\cal K}(G)$ are defined.
We assume w.l.o.g.\ that $G = G_0 \cup G_1$, 
where $G_0$ contains no function symbols in 
$\Sigma$ and $G_1$ consists of ground unit clauses of the form
$f(c_1, \dots, c_n) \approx c$
where 
$c_i, c$ are constants in $\Sigma_0 \cup \Sigma_c$
and $f \in \Sigma$.

\smallskip
\noindent We construct another structure,  $\A$, having the same support as $P$, 
which inherits all relations in ${\sf Pred}$ and 
all maps in $\Sigma_0 \cup \Sigma_c$ from $P$, but on which  
the domains of definition of the $\Sigma$-functions are restricted 
as follows: for every $f \in \Sigma$, 
$f_{\A}(a_1, \dots, a_n)$ is defined if and only if 
there exist constants $c^1, \dots, c^n$ such that 
$f(c^1, \dots, c^n)$ is in $\Psi_{\cal K}(G)$ and 
$a^i = c^i_P$ for all $i \in \{ 1, \dots, n \}$. 
In this case we define $f_{\A}(a_1, \dots, a_n) := f_P(c^1_P, \dots, c^n_P)$.
The reduct of $\A$ to $(\Sigma_0 \cup \Sigma_c, {\sf Pred})$ 
coincides with that of $P$. 
Thus, $\A$ is a model of ${\mathcal T}_0 \cup G_0$. 
By the way the operations in $\Sigma$ are defined in $\A$ it is 
clear that $\A$ satisfies $G_1$, so $\A$ satisfies $G$. 

\smallskip
\noindent We now show that $\A \models_w {\mathcal K}$.
Let $D$ be a clause in ${\mathcal K}$. 
If $D$ is ground then all its terms are defined, and all terms
starting with an extension function are contained in $\Psi_{\K}(G)$, i.e.\ 
$D \in {\mathcal K}[\Psi_{\K}(G)]$, so $D$ 
is true in $P$, hence it is also true in $\A$.

\smallskip
\noindent Now consider the case in which $D$ is not ground. 
Let $\beta : X \rightarrow \A$ be an arbitrary
valuation. Again, if there is a term $t$ in $D$ such that $\beta(t)$ is undefined, we
immediately have that $\beta$ weakly satisfies $D$. 
So let us suppose that for all terms $t$ occurring in $D$, $\beta(t)$
is defined. We associate with $\beta$ a substitution $\sigma$ as follows: 
Let $x$ be a variable.We have the following possibilities: 

\smallskip
\noindent {\em Case 1:} $x$ does not occur below any extension
function. This case is unproblematic. We can define $\sigma(x)$ arbitrarily.

\smallskip
\noindent {\em Case 2:}  $x$ occurs in a unique term $t = f(...x...y...)$ 
(which may occur more than once) and $x$ occurs only once in $t$. 
From the fact that $\beta(t)$ is 
defined, we know that
there are ground terms which we will denote by $t_x, t_y, \dots$ such
that $\beta(x) =(t_x)_P, \beta(y) =(t_y)_P, \dots$. 
Since $\beta(t) =
f_A(...(t_x)_P \dots (t_y)_P \dots)$ is defined,  $f(\dots, t_x, \dots, t_y,
\dots) \in \Psi_{\cal K}(G)$. 
We can define $\sigma(x) = t_x$. 

\smallskip
\noindent {\em Case 3:}  $x$ occurs in two or more terms of the form $f_k(x^k_1,
\dots, x, \dots,  x^k_{n_k})$, $1 \leq k \leq p$, $p \geq 2$, but occurs at most once in
any term of $C$, where $f_1, \dots, f_n$ are function symbols, not
necessarily different (but in terms starting with the same function symbols
$x$ occurs on different positions).

\noindent From the fact that $\beta(f_k(x^k_1, \dots, x, \dots, x^k_n))$ is  
defined, we know that
there are ground terms which we will denote by $t^k_{x}, t_{x^k_i}$ such
that for every $k$ with $1 \leq k \leq p$:
\begin{itemize}
\item $\beta(x) =(t^k_x)_P, \beta(x^k_i) =(t_{x^k_i})_P$ for $1 \leq k
\leq p$ and $1 \leq i \leq n_k$, and 
\item $\beta(f_k(x^k_1, \dots, x, \dots, x^k_{n_k})) =
  f_{\A}((t_{x^k_1})_P, ..., (t^k_x)_P,  \dots,  (t_{x^k_n})_P)$,\\
 i.e.\  $f(t_{x^k_1}, \dots, t^k_x, \dots, t_{x^k_n}) \in \Psi_{\cal K}(G)$.
\end{itemize}
We know that $\Psi_{\K}$ has the property that for every 
clause $C \in \K$, if $C$ contains terms $f_i(x^i_1, \dots, x, \dots,
x^i_{n_i}) $ and $f_k(x^k_1, \dots, x, \dots, x^k_{n_k})$ and if 

$f_i(t_{x^i_1}, \dots, t^i_x, \dots, t_{x^i_{n_i}})    \in \Psi_{\cal
  K}(G)$ and 
$f_k(t_{x^k_1}, \dots, t^k_x, \dots, t_{x^k_{n_k}}) \in \Psi_{\cal K}(G)$ 

\noindent then also $f_i(t^i_{x^i_1}, \dots, t^k_x, \dots, s_{x^i_{n_i}})  \in \Psi_{\cal
  K}(G)$
and$f_k(t^k_{x^k_1}, \dots, t^i_x, \dots, s_{x^k_{n_k}})  \in
\Psi_{\cal K}(G)$. 

This means that we can define 
$\sigma(y) = t_y$ for every linear variable; 
for every variable $x$ which occurs in different terms, let $t_x$ be
one of the terms obtained as before (say $t_x = t^1_x$) and define $\sigma(x) = t_x$. 

\smallskip
\noindent 
Thus, we can construct a substitution $\sigma$ with 
$\sigma(D) \in {\mathcal K}[G]$ and $\beta \circ \sigma = \beta$. 
As $(P, \beta) \models_w \sigma(D)$ we can infer $(\A, \beta) \models_w D$.

\smallskip
\noindent 
We now show that 
$D(\A) = \{ f(a_1, \dots, a_n) \mid f_A(a_1, \dots, a_n) \text{ defined} \}$ is closed under $\Psi_{\cal K}$. 
By definition, $f(a_1, \dots, a_n) \in D(\A)$ iff  there exist 
$\text{ constants } c_1, \dots, c_n$ with ${c_i}_A = a_i$ for all $i$ and $f(c_1, \dots, c_n) \in \Psi_{\cal K}(G)$. Thus, 
$$\begin{array}{rll}
D(\A) & =  \{ f(a_1, \dots, a_n) \mid f_A(a_1, \dots, a_n) \text{ defined} \} & \\
      & =  \{ f({c_1}_{\A}, \dots, {c_n}_{\A}) \mid c_i \text{ constants with } f(c_1, \dots, c_n) \in \Psi_{\cal K}(G) \} & \\
      & =  {\overline h}(\Psi_{\cal K}(G)) & \!\!\!\!\!\!\!\!\!\!\!\!\!\!\!\!\!\!\!\!\!\!\!\!\!\!\!\!\!\!\!\!\!\!\!\!\!\!\!\!\!\!\!\!\!\!\!\!\!\!\!\!\!\!\!\!\!\!\!\!\!\!\text{ where } h(c_i) = a_i \text{ for all } i \\
\Psi_{\cal K}(D(\A)) & =  \Psi_{\cal K}({\overline h}(\Psi_{\cal K}(G)))  = {\overline h}(\Psi_{\cal K}(\Psi_{\cal K}(G))) & \!\!\!\!\!\!\!\!\!\!\!\!\!\!\!\!\!\!\!\!\!\!\!\!\!\!\!\!\!\!\!\!\!\!\!\!\!\!\!\!\!\!\!\!\!\!\!\!\!\!\!\!\!\!\!\!\!\!\!\!\!\!\text{ by property (iv) of } \Psi \\
& \subseteq {\overline h}(\Psi_{\cal K}(G)) = D(\A) & \!\!\!\!\!\!\!\!\!\!\!\!\!\!\!\!\!\!\!\!\!\!\!\!\!\!\!\!\!\!\!\!\!\!\!\!\!\!\!\!\!\!\!\!\!\!\!\!\!\!\!\!\!\!\!\!\!\!\!\!\!\!\text{ by property (iii) of } \Psi\\
\end{array}$$

\noindent As $\A \models_w {\mathcal K}$, 
$\A$ weakly embeds into a total algebra $\B$ satisfying 
${\mathcal T}_0 \cup {\mathcal K}$. 
But then $\B \models G$, 
so $\B \models {\mathcal T}_0 \cup {\mathcal K} \cup G$, 
which is a contradiction. \QED

\

\noindent 
{\bf Remark:} A similar result can be proved also in the case in which 
some variables occur several times below a function symbol
if $\Psi_{\K}$ has the property that 
if $f(x_1, \dots, x, \dots, x \dots x_n) \in \K$
and $f(t_1, \dots, s, \dots, t, \dots, t_n) \in \Psi_{\K}(T)$\\
then $f(t_1, \dots, t, \dots, t, \dots, t_n) \in \Psi_{\K}(T)$
and $f(t_1, \dots, s, \dots, s, \dots, t_n) \in \Psi_{\K}(T)$.

\section{Constrained Horn Clauses: Definitions}
\label{app:constrained-Horn-Clauses}
We give the definitions of constrained Horn clauses, mainly 
following the presentation in \cite{McMillanRybalchenkoBjorner}.

\begin{defi}[\cite{McMillanRybalchenkoBjorner}]
Conjunctions $\Pi$ of constrained Horn clauses  are constructed as follows:
\begin{align*}
\Pi & ::= {\sf chc} \wedge \Pi \mid \top\\
{\sf chc} & ::= \forall var . {\sf chc}  \mid {\sf body} \rightarrow {\sf  head} \\
{\sf pred} & ::= {\sf upred} \mid \phi \\
{\sf head} &  ::= {\sf pred} \\
{\sf body} & ::= \top \mid {\sf pred} \mid  {\sf body} \wedge {\sf body} \mid
\exists var . {\sf body} \\
{\sf upred} &  ::= \text{an uninterpreted predicate applied to terms}\\
\phi & ::= \text{a formula whose terms and predicates are
  interpreted over } \A \\
var & ::= \text{a variable}
\end{align*}

\noindent A clause where the head is a formula $\phi$ is called a query or a
goal clause. 
The terminology ``fact clause'' is used for a clause whose head is an uninterpreted
predicate and body is a formula $\phi$.
\end{defi}
It is easy to see that in Theorem~\ref{thm:min-model}, if we guarantee that the formulae
$\phi_i$ are formulae whose terms and predicates are interpreted over $\A$ 
then all clauses of the form 
$$\phi_i({\overline x}) \rightarrow  \mu_i({\overline x})$$
are constrained Horn clauses, hence: 

\medskip
$\begin{array}{rl} 
CH_N = \{ & \phi_i({\overline x}) \rightarrow  \mu_i({\overline x})
 \mid  i \in \{
1, \dots, n \}\} \\
\cup \{& (\mu_i({\overline x}) \wedge \mu_j({\overline y}))\sigma \rightarrow
\mu_k({\overline z})  \mid C_k({\overline z}) \text{ is obtained by a
    resolution }  \\
& \hspace{1.5cm} \text{ inference in } {\cal I}_P\text{ from } C_i({\overline x}) \text{ and } C_j({\overline y}) \text{ with m.g.u.\ } \sigma \}\\
\cup \{& \mu_i({\overline x})\sigma  \rightarrow \mu_k({\overline z})
\mid  C_k({\overline z}) \text{ is obtained by a 
    factorization inference  }  \\
&  \hspace{1.5cm} \text{ in } {\cal I}_P \text{ from } C_i({\overline x}) \text{ with m.g.u.\ } \sigma \}
\end{array}$

\medskip
\noindent is a set of constrained Horn clauses. 
\end{document}